\documentclass[journal ]{new-aiaa}
\usepackage[utf8]{inputenc}
\usepackage{textcomp}
\usepackage{rotating}
\usepackage{algorithm}
\usepackage{algpseudocode}
\usepackage{graphicx}
\usepackage{amsmath}
\usepackage[version=4]{mhchem}
\usepackage{siunitx}
\usepackage{longtable,tabularx}
\usepackage{cleveref}
\usepackage{subcaption}
\usepackage{tikz}
\usepackage{booktabs}
\usepackage{todonotes}
\crefname{algorithm}{Algorithm}{Algorithms}
\Crefname{algorithm}{Algorithm}{Algorithms}

\definecolor{darkblue}{RGB}{0,32,96}
\definecolor{deepblue}{RGB}{27,136,185}
\definecolor{lightblue}{RGB}{214,239,250}

\definecolor{myRed}{RGB}{178,34,34}
\definecolor{myBlack}{RGB}{36,36,36}
\definecolor{myOrangeRed}{RGB}{233,80,45}
\definecolor{myOrange}{RGB}{233,143,44}

\title{Critical slowing down for predicting controller induced loss of control in quadrotors}

\author{Jasper J. van Beers\footnote{PhD Candidate, Faculty of Aerospace Engineering, Delft University of Technology, 2629 HS Delft, The Netherlands}, Prashant Solanki \footnote{PostDoc, Faculty of Aerospace Engineering, Delft University of Technology, 2629 HS Delft, The Netherlands}, Erik-Jan van Kampen \footnote{Associate professor, Faculty of Aerospace Engineering, Delft University of Technology, 2629 HS Delft, The Netherlands}, and Coen C. de Visser \footnote{Associate professor, Faculty of Aerospace Engineering, Delft University of Technology, 2629 HS Delft, The Netherlands}}
\affil{Delft University of Technology, 2629 HS Delft, The Netherlands}

\begin{document}

\maketitle

\begin{abstract}
We develop a novel forecasting scheme to anticipate controller induced loss of control (LOC) in quadrotors and evaluate it on real LOC flight data from four different quadrotors. For this, early warning signals of LOC are derived using critical slowing down (CSD), a generic phenomenon shown to precede critical transitions across various complex ecological and biological systems. As such, our early warning indicators are generic in the sense that no system models are needed to facilitate forecasts of LOC. The approach is evaluated on real quadrotor flight data wherein LOC occurs due to unstable controller behavior arising from input-output delays. Our approach achieves a time-to-LOC forecast of up to 0.9 seconds before LOC occurs, outperforming state-of-the-art recurrent neural network quadrotor LOC forecasters in terms of detection accuracy and LOC data reliance. In particular, we leverage insights from CSD to accurately predict LOC without using data of the LOC event itself. Going further, we apply our forecasters without any re-parameterization to anticipate a different LOC scenario, quadrotor flyways, that occur on other quadrotors flying both indoors and outdoors. Despite these differences, our approach successfully detects LOC, demonstrating that it can generalize across controller architectures, quadrotors, and LOC scenarios.
\end{abstract}

\section{Introduction}
\lettrine{A}{s} autonomous robots become more prevalent in both industry and society, ensuring their safe operation is paramount. This acute need to establish safety is affirmed by studies which find that increasing automation can often provoke more accidents \cite{OLIVER2019772_acLOC,YANG2022105623_robotsafer,RobotFailure2004}. Many of these arise from loss of control (LOC) \cite{RobotFailure2004}, wherein the closed-loop system becomes unable to maintain the desired behavior.

These concerns of safety are particularly relevant for many unmanned aerial vehicles (UAVs) \cite{app12031047,drones6060137,nisingizwe2022effect}. A particularly popular UAV in safety research is the under-actuated quadrotor, for which a considerable body of literature addresses fault detection and fault-tolerant control (FTC). As such, many impressive advances have been made in recent years on both partial \cite{Guo2022_Quad_GE_blade_damage,Eltrabyly2022_QuadFTC,Madruga2023_LOE_UAV_actuators} and full rotor losses \cite{Ke_2023_Quadrotor_123FTC,Sun2020_UpsetRecovery,Nan2022_NMPC_FTC,Sun_ControlDoubleFailure}, allowing quadrotors to maintain rudimentary control despite (multiple) rotor failures. However, even a fully rotor functional quadrotor may still lose control: for example, due to sensor saturation or unstable controller characteristics that lead to `flyaways' \cite{drones9020141_flyaways}. 

Therefore, a growing body of research instead addresses the issue of safety by defining and enforcing safe operating spaces for dynamical systems. For quadrotors, this is often achieved by constraining it's position or attitude. Subsequently, control barrier functions (CBFs) are often applied to ensure that the system remains within the prescribed bounds (e.g., \cite{Gurriet2020_SCCF_nonlinearsys,Zheng2023_Geofencing_Quad,Singletary2022_OnboardSafety_RacingQuad_GeofencingCBF}). For instance, Singletary et al. \cite{Singletary2022_OnboardSafety_RacingQuad_GeofencingCBF} use CBFs to allow high-speed (up to 28m/s) quadrotor flight outdoors within the bounds of a pre-defined 3-dimensional positional geo-fence. Updating these constrained regions online also enables quadrotors to avoid collisions with moving and unpredictable agents \cite{Fridovich-Keil2020_CollisionAvoidance,Singletary2022_SafeDroneFlight_TBC,Li2024}. 
However, these approaches effectively bound navigational states and do not constrain the inner-loop state and control variables, such as the body and motor rotational rates. Therefore, disturbances may still upset these unrestricted states in a way that, left unchecked, propels the system into an unrecoverable state. 

Such events are preventable through knowledge of the quadrotor's control invariant set \cite{Sun2019_MonteCarlo}, or safe flight envelope (SFE) in aerospace applications. In this context, LOC is often associated with an excursion beyond the SFE bounds \cite{belcastro2012loss}. While SFEs are well-studied for aircraft \cite{Barlow2011_EstimatingLOC_DataDriven,Rafi2021_Realtime_LOC_mitigation,Zhidong2022_FEP_Reachability}, there is limited work available on quadrotors. Unlike geo-fenced regions, high-dimensional reachable sets are computationally expensive to determine through standard level-set methods \cite{Sun2019_MonteCarlo,Zhidong2022_FEP_Reachability,bansal2017hamilton}. Moreover, any steady-state simplifications made to reduce the SFE dimensionality for fixed-wing aircraft (e.g., \cite{Zogopoulos2021_FTC_FixedWingUAV_FE_awareness}) are not always possible for the quadrotor. Thus, Sun et al. \cite{Sun2019_MonteCarlo} instead approximate a reduced order 6-state SFE of a quadrotor through Monte-Carlo simulations of the forward and backward reachable sets about an equilibrium condition, subject to a bang-bang optimal controller. The resultant SFE is inherently model and controller dependent. Such a scheme alone is unsuitable for cases where the system dynamics change unexpectedly. Even if the dynamics are unchanging, model mismatches nonetheless propagate into the SFE. 

A complementary view to the global bounds offered by SFEs is to consider the local proximity of the quadrotor to LOC. For instance, Altena et al. \cite{Altena_ANN_LOC} apply neural networks to detect and predict LOC in small quadrotors using only on-board sensor data. While LOC could be successfully predicted, the resultant networks exhibited unsatisfactory generalization capabilities. Moreover, the black-box nature of the neural networks obscures their physical interpretability. To facilitate such explainability, useful early warning signals (EWS) which act as precursors to quadrotor LOC need to be identified. 

To address this, we draw inspiration from the field of ecology and apply critical slowing down (CSD) \cite{Scheffer2009} to predict LOC in quadrotors. CSD is a generic phenomenon that describes the slowing recovery rate dynamics that a system experiences when approaching a critical transition \cite{Scheffer2009}. It has been shown to anticipate critical transitions across a plethora of complex natural systems, from the onset of human seizures \cite{Maturana2020} to disease (re-)emergence \cite{tredennick2022anticipating_measles,Brett2020}. This also includes demonstrations on real world measurement data, such as for predicting the onset of traffic congestion \cite{Ghadami2021_trafficCongestion,Ghadami2022_TrafficJam}; detecting haptic instabilities \cite{kerr2023haptic}; and forecasting the collapse of key ocean current systems \cite{ditlevsen2023warning_AMOC_CSD}. Moreover, it has recently been shown that CSD is also effective at monitoring loss of stability in controlled systems like the quadrotor \cite{vanBeers2026_ews_for_loc}.

Building on this, the main contribution of this paper is a system model-free forecasting framework for anticipating controller induced (i.e., closed-loop unstable) loss of control (LOC) in quadrotors. The approach uses a series of early detections of LOC, observed through generic CSD indicators, to formulate time-to-LOC \textit{forecasts} of up to 0.9 seconds before the LOC event occurs. For a fast system like the quadrotor, this offers ample forewarning to the controller such that LOC can be prevented in time. Furthermore, the approach relies only on commonly available onboard sensor information and is validated on real world flight data of four different quadrotors, including 91 actual LOC events.

We show that the our approach outperforms existing LOC forecasters from \cite{Altena_ANN_LOC} in terms of LOC classification accuracy and LOC data reliance, which is a key advantage over many data-driven methods. By leveraging the generic principles of CSD, an initial estimate of LOC behavior can be trivially made \textit{without} actual LOC data. This also allows the C-BeFore to abstract LOC detection across different LOC scenarios and (quadrotor) platforms. We stress that our approach is not a controller, but rather an `add-on' safety monitor for LOC that informs a controller of upcoming LOC events such that corrective action can be taken. That is, we propose an additional layer of safety \textit{independent} of the controller which can then be used in tandem with existing (e.g., fault tolerant) controllers.

The remainder of the paper is organized as follows: \cref{sec:background} introduces critical slowing down theory. Next, \cref{sec:quadLOC} presents examples of controller-induced quadrotor LOC. The early warning signal and LOC forecaster design are then explained in \cref{sec:forecaster_design}. \Cref{sec:results} presents the forecasting results for real quadrotor LOC data along with some limitations of the approach. The paper concludes with \cref{sec:conclusion}.

\section{PRELIMINARIES}\label{sec:background}

\subsection{Critical slowing down}
The core principle behind critical slowing down (CSD) is that, as a complex system approaches a critical transition, it becomes less responsive to perturbations about its equilibrium \cite{Scheffer2009}. Typically, the system's resilience to these perturbations also deteriorates in tandem \cite{forzieri2022emerging}. 

\begin{figure}[!t]
    \centering
    \includegraphics[width = \columnwidth]{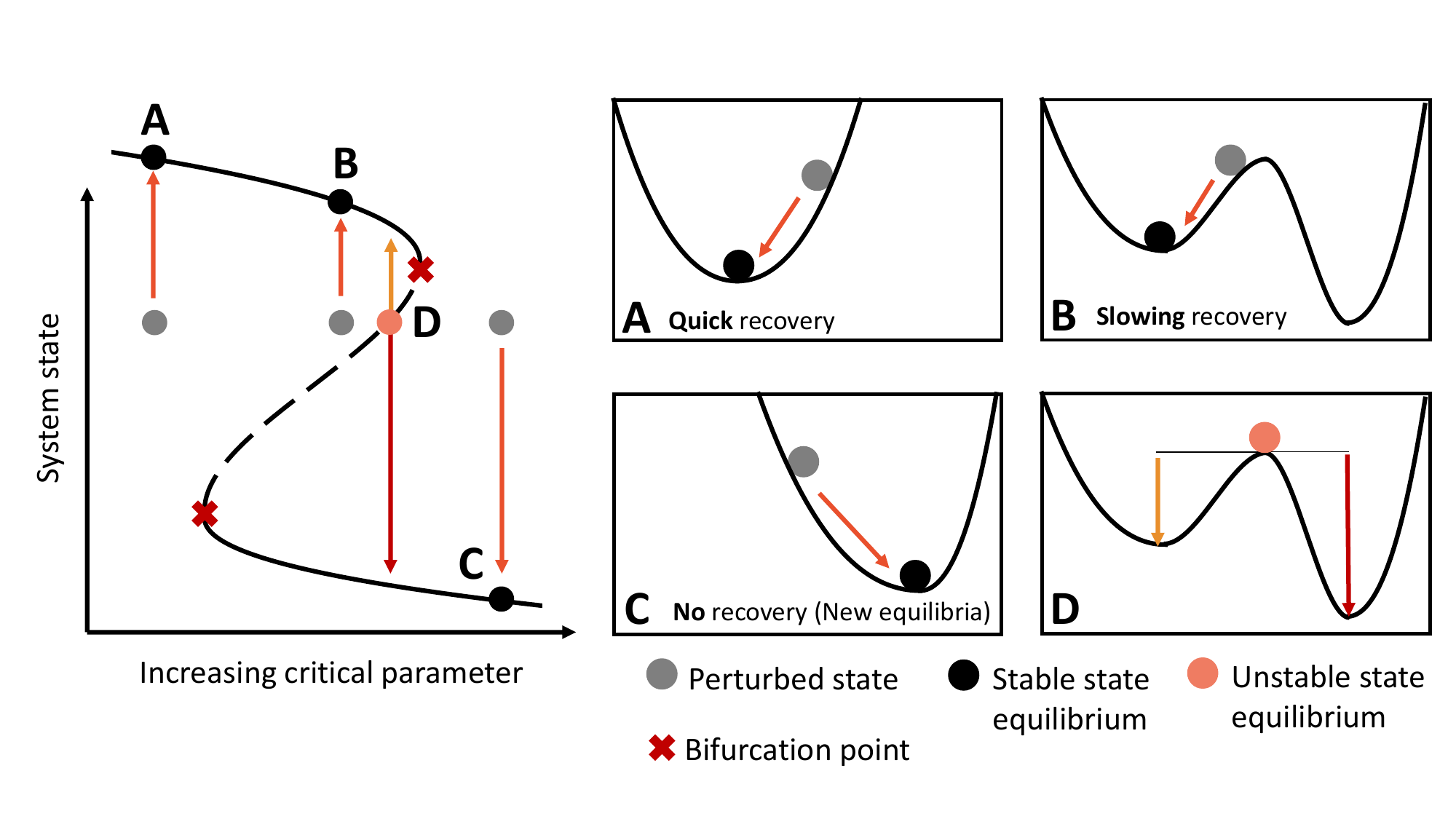}
    \caption{Illustration of the critical slowing down (CSD) phenomenon through a fold bifurcation (left plot) at various equilibrium points (\textbf{A}, \textbf{B}, \textbf{C}, \textbf{D}) with transition points denoted by the red crosses. As a system approaches a critical transition, with increasing critical parameter (x-axis) from \textbf{A} to \textbf{D}, it becomes slower to recover from small perturbations (gray ball) about its stable equilibrium point (black ball).} 
    \label{fig:CSD_diagram}
\end{figure}

This relation between equilibria, responsiveness, and resilience is intuitively explained through a system experiencing a fold (or saddle-node) bifurcation, although the fundamental phenomena is analogous across other bifurcations which exhibit CSD. Fig. \ref{fig:CSD_diagram} illustrates a fold bifurcation (left plot) governed by a critical parameter (x-axis) with tipping points marked by the red crosses. Far from these points, the rate of recovery of the system from perturbations is strong (e.g., \textbf{A} and \textbf{C}). However, as a bifurcation is approached, the system becomes slower to respond to perturbations due to contracting regions of attraction (compare recovery rate of \textbf{A} with \textbf{B}). In some cases, the system becomes less resilient; disturbances previously accommodated may now induce a transition by pushing the system over the shrinking `hump' (i.e., unstable equilibrium) dividing the equilibria (\textbf{D}).

These characteristics of CSD are consistent with the change in dynamic behavior experienced by a control system as it transitions from stability to instability \cite{vanBeers2026_ews_for_loc}. This link is intuitive for linear control systems: the characteristic slowing recovery rate of CSD implies that the real part of the system's dominant eigenvalues approach zero (i.e., instability). Nonetheless, the implications on stability of such slowing down extends also to nonlinear (control) systems (e.g., \cite{Scheffer2009,vanBeers2026_ews_for_loc,forzieri2022emerging}), making CSD especially suitable for monitoring loss of control. 

CSD is typically measured through two metrics: the lag-1 autocorrelation\footnote{That is, the correlation of a signal with itself lagged by one sample.} (AC1) and variance (SD). These indicators are measured along a running window applied to a detrended\footnote{As CSD concerns a systems response to perturbations about some equilibrium, it is important to first remove any slow trends from the signal.} signal of interest. As a critical transition nears, it is expected that $\text{AC1} \to 1$ and $\text{SD} \to \infty$ \cite{Scheffer2009,Scheffer_2008_CriticalTransitions,Dakos2012}. In this paper, we favor the AC1 indicator due to its boundedness (i.e., AC1 $\in$ [-1, 1]) and predictable behavior when nearing a transition (i.e., AC1 $\to 1$). These predictable properties limit the need for LOC data, which is especially useful in safety applications where such data is often difficult to reliably and safely obtain.

\section{QUADROTOR LOSS OF CONTROL}\label{sec:quadLOC}

\begin{figure*}[!t]
    \centering
    \includegraphics[width=\linewidth]{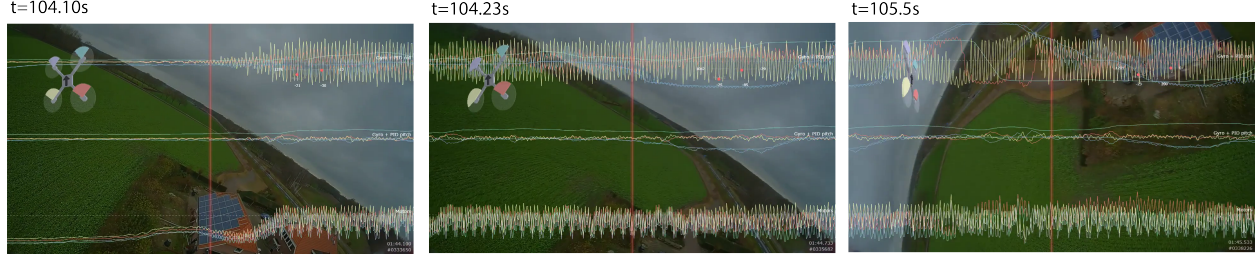}
    \caption{Example of a flyaway event that occurs on a human-piloted FPV quadrotor. The measured rotors speeds (top), gyroscope rates (middle), and low-level Betaflight controller motor commands (bottom) are overlaid on the FPV footage.}
    \label{fig:flyaway_sk5}
\end{figure*}
Critical slowing down can be observed in control systems that, either continuously or incrementally, \textit{approach} loss of control (LOC). Examples of these for quadrotors include loss of control due to integrator windup, input-output delays, sensor saturation, structural resonance, changes to system dynamics (e.g., due to the build up of faults), and unmodeled dynamics including interactions with attached loads/manipulators or the environment. Contrarily, LOC events which are effectively instantaneous, such as quadrotor collisions, impacts, and sudden catastrophic faults, are out of scope and various methods that seek to accommodate these problems are addressed in existing literature \cite{Sun2020_UpsetRecovery,Nan2022_NMPC_FTC,Sun_ControlDoubleFailure,Fridovich-Keil2020_CollisionAvoidance}.

Instead, we focus on the less common - yet equally as catastrophic - LOC events that occur on nominal fault-free quadrotors. An infamous example of this is the quadrotor `flyaway' (i.e., rapid uncontrolled ascent) induced by severe oscillations in the rotor speed commands \cite{drones9020141_flyaways,vanBeers2026_ews_for_loc}. Such flyaways can arise from poor controller characteristics at certain regions of the flight envelope which can remain elusive until after they occur. For example, structural resonance can provoke flyaways in hobbyist FPV drones, as seen in during outdoor human piloted flight testing of one such quadrotor (\cref{fig:flyaway_sk5}). Similarly, poor controller characteristics can trigger flyaway events following strong disturbances, such as wind gusts. While the underlying controllers can be tuned to mitigate these effects after the instability is made apparent, other such controller vulnerabilities to LOC can emerge for the updated controller or remain hidden altogether. As such, predicting when these flyaway events will occur is an outstanding challenge. 

\subsection{Benchmark loss of control data set}

One of the main issues involved with preventing quadrotor flyaway events is that they are difficult to predict. On the other hand, these flyaway events fundamentally arise from unstable controller behavior driven, in part, by the current system state. Such a loss of control (LOC) scenario is studied in \cite{Altena_ANN_LOC}, for which an extensive LOC data set was collected. Specifically, in \cite{Altena_ANN_LOC}, quadrotors are driven into LOC through a high yaw rate ($\pm 2000$ deg/s) maneuver. Due to the coupling of this rapid yaw rotation and the inherent motor input-output delay of the actuators, unstable off-axis (i.e., roll and pitch) oscillations grow and eventually culminate in a crash (see \cref{fig:cinego_loc_snapshot}). While this yaw-induced LOC scenario seldom occurs in real quadrotor flight (unless a complete rotor fault occurs \cite{Ke_2023_Quadrotor_123FTC,Nan2022_NMPC_FTC,Sun_ControlDoubleFailure}), it serves as a consistent proxy for controller-induced instabilities upon which LOC forecasters can be evaluated. 

To this end, we employ the same LOC labeling scheme as \cite{Altena_ANN_LOC}: LOC is defined as the moment in which roll ($\phi$) or pitch ($\theta$) first exceeds $\pm 90$ degrees (see \cref{eq:LM-ATT}). This definition is chosen since the quadrotor's attitude loop controller\footnote{The experiments were flown using Betaflight's horizon mode.} assumes $|\phi|, |\theta| < 90$. While this is arguably an extreme condition for the quadrotor, the goal is to anticipate this moment \textit{before} it actually occurs.

\begin{equation}\label{eq:LM-ATT}
    \text{LM-ATT} = \left \{ \begin{array}{cc}
        0 & \text{if } |\phi| > 90  \text{ or } |\theta| > 90 \\
        1 & \text{otherwise} 
    \end{array} \right .
\end{equation}

Note that our approach (and CSD in general \cite{vanBeers2026_ews_for_loc}) is \textit{not} restricted to this aggressive yawing maneuver. It is the underlying closed-loop instability - induced here by input-output delay - which concerns our approach. To this end, we also apply our approach to anticipate the aforementioned flyaway events.

\noindent
\begin{minipage}[t]{0.43\textwidth}
    \centering
    \includegraphics[width=\linewidth]{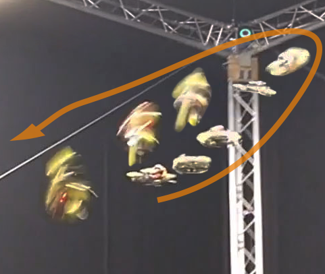}
    \captionof{figure}{Composite image depicting the evolution of the yaw induced loss of control scenario.}
    \label{fig:cinego_loc_snapshot}
\end{minipage}
\hfill
\begin{minipage}[t]{0.55\textwidth}
    \centering
    \includegraphics[width=\linewidth]{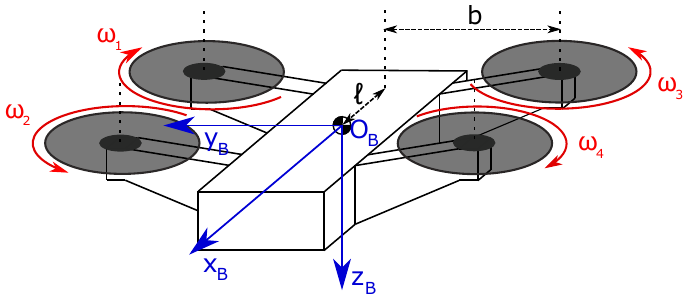}
    \captionof{figure}{Quadrotor body reference frame denoted by $x_{B}$, $y_{B}$ and $z_{B}$.}
    \label{fig:quadrotor_Bframe}
\end{minipage}

\section{FORECASTERS FOR QUADROTOR LOSS OF CONTROL}\label{sec:forecaster_design}

\subsection{Early warning signals}\label{subsec:EWS_design}

\subsubsection{Quadrotor variables}

In line with earlier studies on quadrotor loss of control (LOC) \cite{Altena_ANN_LOC,vanBeers2026_ews_for_loc,vanBeers_FCM}, we use the onboard measurements of the rotor speeds ($\omega_{1}, \omega_{2}, \omega_{3}, \omega_{4}$) as the basis for our early warning indicators. These measurements reflect the inputs of the controller and are thus sensitive to instabilities that emanate from the controller's actions. To relate these motor outputs to the rotational dynamics of the quadrotor, we sum them about each of the quadrotor's axes\footnote{Following the layout of \cref{fig:quadrotor_Bframe}.}:
\begin{equation}\label{eq:rotor_diff}
    \begin{array}{cl}
        u_{p} = & \left ( {\omega_{3}} + {\omega_{4}} \right ) - \left ( {\omega_{1}} + {\omega_{2}} \right ) \\
        u_{q} = & \left ( {\omega_{2}} + {\omega_{4}} \right ) - \left ( {\omega_{1}} + {\omega_{3}} \right ) \\
        u_{r} = & S_{r,1} \left [ \left ( {\omega_{1}} + {\omega_{4}} \right ) - \left ( {\omega_{2}} + {\omega_{3}} \right ) \right ]
    \end{array}
\end{equation}
\noindent
where $S_{r, 1} = 1$ for clockwise rotation of $\omega _{1}$ and $S_{r, 1} = -1$ otherwise. We refer to $u_{p}, u_{q}, u_{r}$ as the rolling, pitching, and yawing rotor speed differences, respectively. 

\subsubsection{Critical slowing down parameters}\label{subsubsec:csd_ews_params}

Slow-moving trends and biases are removed from the measured rotor speed differences through a moving average detrender \cite{Brockwell2016} for simplicity. The degree of detrending is controlled by the moving average window size, $W$, and a detrended sample, $x'[k]$, is calculated through:
\begin{equation*}
    x'[k] = x[k] - \frac{1}{W}\sum_{i=k-W}^{k}x[i]
\end{equation*}

\begin{figure}[!t]
    \centering
    \includegraphics[width=0.8\linewidth]{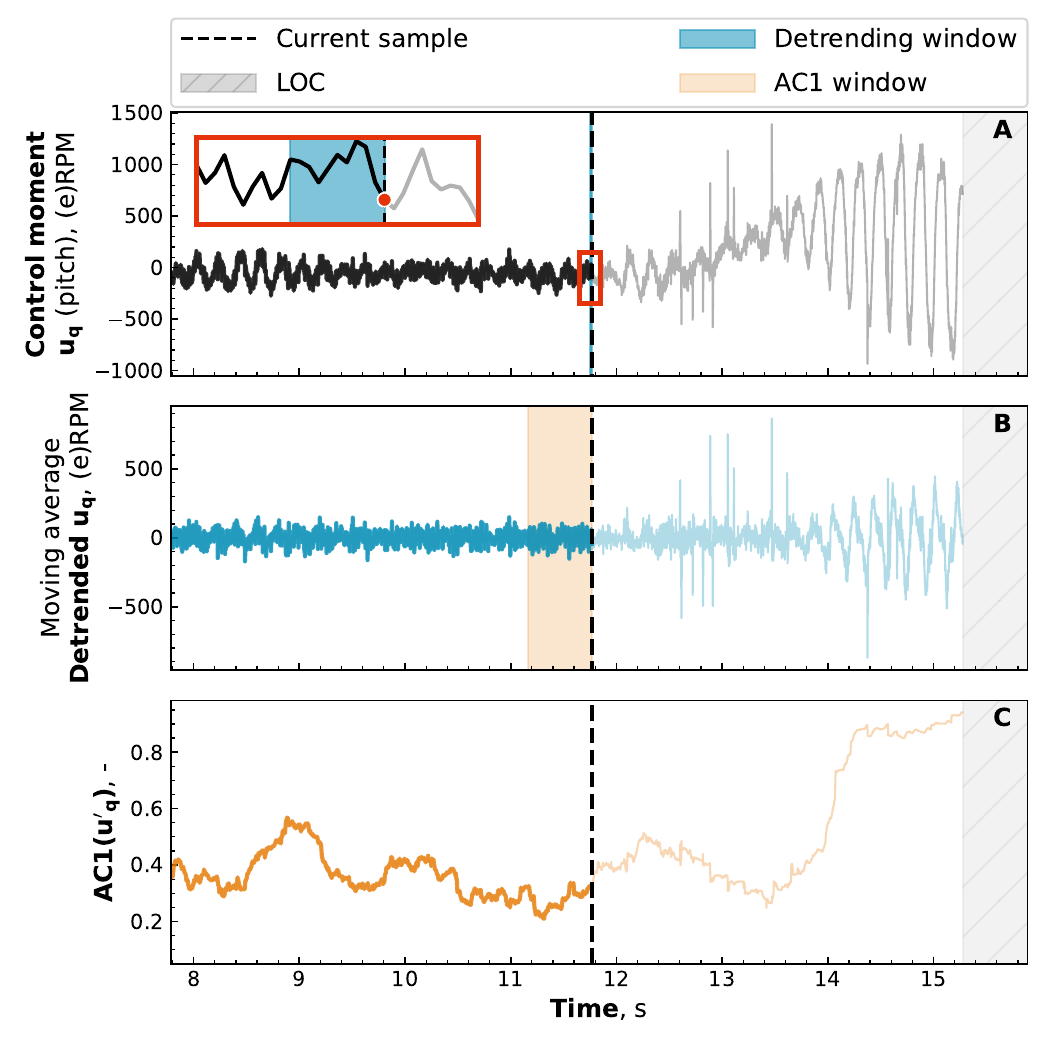}
    \caption{Example of how the lag-1 autocorrelation (AC1) indicator of critical slowing down is constructed for the pitching rotor speed difference, $u_{q}$, on real flight data. The unaltered signal, shown in \textbf{A}, is detrended via a moving average detrender (window size of $10$ samples), shown in \textbf{B}. Then, the AC1 is calculated over a window of 300 samples, shown in \textbf{C}. Loss of control (LOC) is denoted by the shaded gray region.}
    \label{fig:FID0084_Example_EWS_Construction_D-MA-50_V-uq}
\end{figure}

Following the detrending, the lag-1 autocorrelation (AC1) is calculated through the Pearson correlation coefficient between the detrended signal and itself lagged by one sample, taken over sliding window, $W_{AC}$. This completes the early warning signal (EWS) construction process. In total, this process has linear runtime complexity, $O(N)$, with respect to a time-series signal of $N$ samples. As an example, \cref{fig:FID0084_Example_EWS_Construction_D-MA-50_V-uq} illustrates how the EWS is computed for the pitching rotor speed difference, $u_{q}$, on real flight data. The characteristic growth in AC1 \cite{Dakos2012} as loss of control is approached is visible, suggesting that CSD is indeed observable in the studied LOC scenario. 

\subsubsection{Early warning signal behavior}\label{subsubsec:ews_construction}

Let $E_{k}$ denote an EWS constructed using a set of parameters $k \in \mathcal{K}$, where $\mathcal{K}$ contains all variations in quadrotor variable, detrending window, and AC1 window. Using nominal flight data, a distribution of $E_{k}$ representing nominal system behavior is constructed, which we denote by $N_{E_{k}}$. If LOC data is available, we may likewise construct a distribution for LOC behavior, $L_{E_{k}}$, from data points of $E_{k}$ at the moment of LOC. However, if LOC data is unavailable, we may instead use insights from CSD to form an assumed LOC distribution, $L^{A}_{E_{k}}$, for example, by taking a normal distribution near $E_{k} = 1$ where tipping is expected to occur following CSD theory. As such, LOC data is not strictly necessary and no training is required for the approach. This is an advantage over many data-driven methods, such as neural networks, which can require extensive LOC data sets.

\subsection{Loss of control forecaster design}
\subsubsection{Bayesian loss of control detector}\label{subsubsec:BID}
The goal of the detector is to recognize LOC through observations of $E_{k}$. These observations are a sequence of the $D$ most recent values of $E_{k}$ as observed from the current sample, $i$:
$$E_{k, D} = \{E_{k}[i-D], E_{k}[i-D+1], ..., E_{k}[i] \}$$

These observations are used alongside CSD theory to motivate a probability of LOC, $P(L)$. It is expected that $E_{k} \to 1$ as LOC is approached \cite{Scheffer2009} since $E_{k}$ is an AC1 metric of CSD. Thus, a reasonable estimate for $P(L)$ is:
$$P(L) = \text{med}(E_{k, D}^{+}), \text{ } E_{k, D}^{+} = \{E_{k, D}[d] : E_{k, D}[d] \geq 0\}$$
\noindent
for all $d \in \{ i-D, i-D+1, ... , i \}$. Here, med$(\cdot)$ denotes the median operator, which is used to limit the effect of outliers within the detection window (e.g., the sudden spikes in \cref{fig:FID0084_Example_EWS_Construction_D-MA-50_V-uq}). Moreover, positive values of $E_{k, D}$ are enforced for valid bounds\footnote{$P(L) \in [0, 1]$ to be a valid probability value. The upper-bound is guaranteed by the upper-bound of $E_{k}$ (i.e., the AC1).} on $P(L)$. However, preliminary tests have shown that this $P(L)$ alone is insufficient for anticipating LOC due to the prevalence of false positive detections. 

Instead, estimates of $P(L)$ may be improved by employing it as a prior in a Bayesian inference formulation. Let this Bayesian LOC detector be defined through \cref{eq:BID}, where $P(L | E_{k, D})$ is the posterior (i.e., improved estimate of $P(L)$). The probability that the current observations, $E_{k, D}$, represent LOC versus nominal behavior is described by $P(E_{k, D} | L)$ and $P(E_{k, D} | \neg L)$ respectively. Furthermore, let $P(\neg L) = 1 - P(L)$. 
\begin{equation}\label{eq:BID}
    P(L | E_{k, D}) = \frac{P(L) P(E_{k, D} | L)}{P(L) P(E_{k, D} | L) + P(\neg L) P(E_{k, D} | \neg L)}
\end{equation}

We propose that $P(E_{k, D} | L)$ and $P(E_{k, D} | \neg L)$ be derived from the similarity between the distribution of $E_{k, D}$ and the corresponding $L_{E_{k}}$ (or $L_{E_{k}}^{A}$) and $N_{E_{k}}$ distributions (see \cref{subsubsec:ews_construction}). The operating assumption is that, if the distribution of $E_{k, D}$ is the same as $L_{E_{k}}$ (or $L_{E_{k}}^{A}$), then the quadrotor is in a state of LOC since $E_{k, D}$ perfectly matches the reference LOC distribution of $E_{k}$.  

To evaluate this degree of similarity, the Wasserstein distance (WD) metric is used. Let $\mathcal{W}^{L}_{k, D}$ denote the WD score between $E_{k, D}$ \& $L_{E_{k}}$ and let $\mathcal{W}^{N}_{k, D}$ describe that between $E_{k, D}$ \& $N_{E_{k}}$. As these are not valid probability values, we instead define a relative similarity score for these distances using \cref{eq:PhiL} and \cref{eq:PhiN}. 
\begin{equation}\label{eq:PhiL}
    \Phi_{L} = 1 - \frac{\mathcal{W}^{L}_{k, D}}{\mathcal{W}^{L}_{k, D} + \mathcal{W}^{N}_{k, D}}
\end{equation}


\begin{equation}\label{eq:PhiN}
    \Phi_{N} = 1 - \frac{\mathcal{W}^{N}_{k, D}}{\mathcal{W}^{L}_{k, D} + \mathcal{W}^{N}_{k, D}}
\end{equation}

These functions have the properties that $\Phi_{L}, \Phi_{N} \in [0, 1]$ and $\Phi_{L} + \Phi_{N} = 1$. Moreover, at each observation sample $i$, we obtain a triple of distributions ($E_{k, D}$, $L_{E_{k}}$, $N_{E_{k}}$) for which there are two possible events: either $E_{k, D}$ tends to $L_{E_{k}}$ or towards $N_{E_{k}}$. The probability assigned to these events occurring are given by $\Phi_{L}$ and $\Phi_{N}$ respectively. For these reasons, we argue that these are suitable probability functions for the binary classification task of LOC studied here.

Furthermore, to mitigate the number of false positives, multiple Bayesian detectors defined using different $E_{k}$ are combined to produce a final LOC detection. For the sake of simplicity, we take the average through \cref{eq:stacked_forecasters}, though other aggregation methods are also possible.
\begin{equation}\label{eq:stacked_forecasters}
    P(L | E) = \frac{1}{K}\sum_{k=1}^{K}P(L | E_{k})
\end{equation}
The Bayesian LOC detector has a runtime computational complexity of $O(D\text{log}(D))$ at inference, where $D$ is the largest detection window size in samples. 

\subsubsection{From detections to forecasts}

Assuming that the Bayesian detector represents the true (labeled) LOC moments well with a window size of $D_{d}$, the forecasting problem simplifies to anticipating when such detections should occur. Specifically, if LOC is indeed being approached, then the observations of $E_{k, D}$ should resemble the LOC distribution, $L_{E_{k}}$, earlier for identical Bayesian detectors operating shorter observation windows $D < D_{d}$ (i.e., as LOC is approached, recent $E_{k}$ should better resemble $L_{E_{k}}$ compared to older $E_{k}$). Crucially, if this trend towards LOC persists, then a series detectors with $D < D_{d}$ should sequentially produce detections of LOC ordered by increasing window size. Then, a temporal forecast of LOC, $\Delta t_{LOC}$, from the current sample $i$ can be computed using the difference in window sizes between these early detections.

\begin{figure}[!t]
    \centering
    \includegraphics[width=\linewidth]{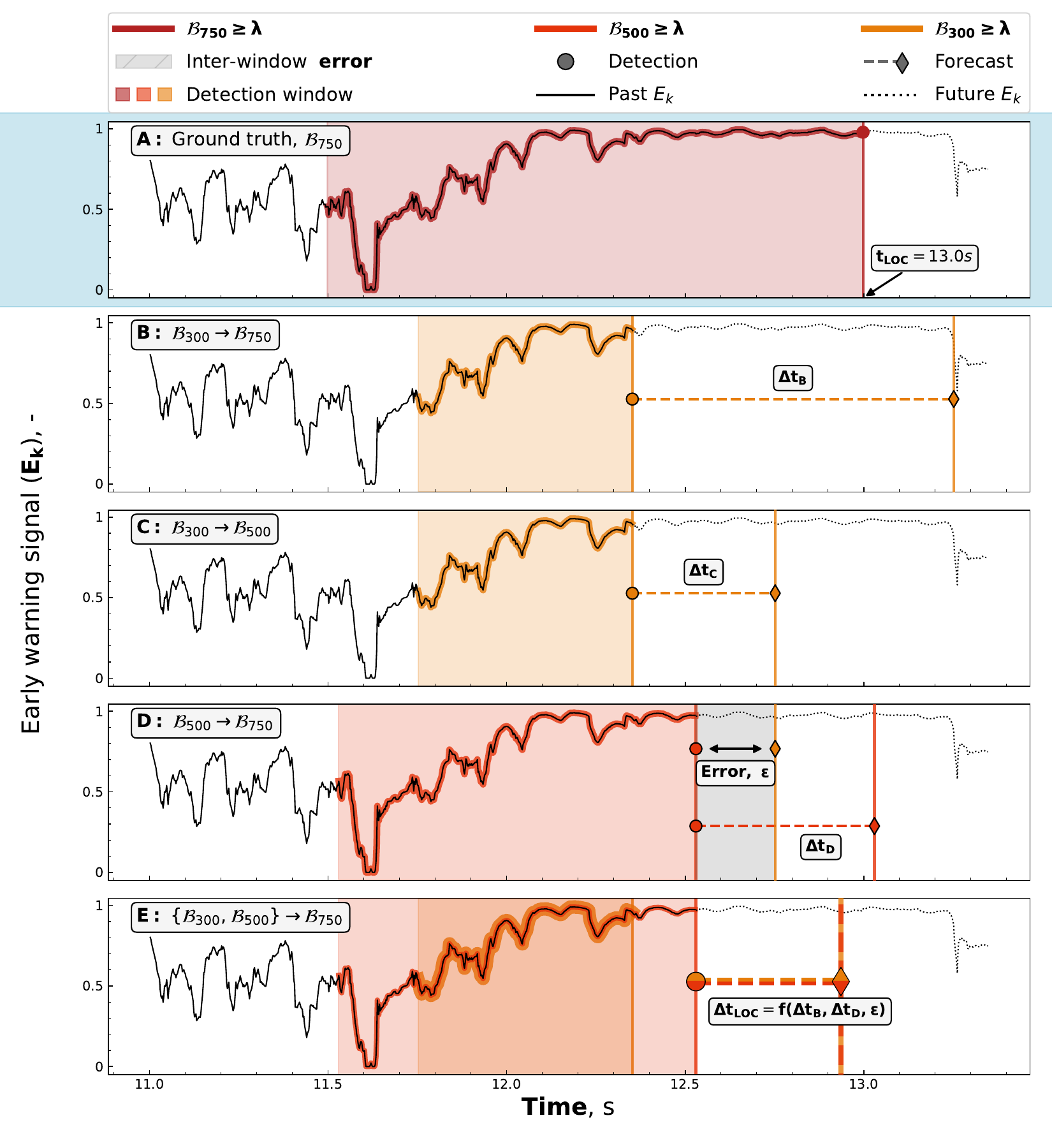}
    \caption{Illustrative example of the loss of control (LOC) forecasting procedure using real LOC flight data. Sequential detections of LOC through shorter window forecasters (subplots B to E) are used to anticipate when the largest window LOC detector (subplot A) is expected to make a detection should the trends in the early warning signal behavior persist.} 
    \label{fig:Example_Time_Forecast}
\end{figure}

This procedure is illustrated through \cref{fig:Example_Time_Forecast} where Bayesian LOC detectors, $\mathcal{B}_{D} = P(L|E_{k, D})$, are applied on real flight data measurements along three different window sizes: $D = \{300, 500, 750\}$ samples. Let $\mathcal{B}_{300}$, $\mathcal{B}_{500}$, and $\mathcal{B}_{750}$ respectively denote these detectors, whose only difference is their window size. It is assumed that a detection made by $\mathcal{B}_{750}$ corresponds to the moment of labeled LOC, $t_{LOC}$, and thus represents the `ground-truth' (highlighted in blue in \textbf{A} of \cref{fig:Example_Time_Forecast}). The goal of the smaller window detectors is to anticipate when $\mathcal{B}_{750}$ should make a detection. For instance, in plot \textbf{B} of \cref{fig:Example_Time_Forecast}, $\mathcal{B}_{300}$ makes such a forecast with $\Delta t_{B} = 450$ (=750-300) samples, assuming that the trend towards LOC persists linearly in time. However, this trend may be accelerating. Thus, $\mathcal{B}_{300}$ is also used to anticipate when $\mathcal{B}_{500}$ should produce a detection (plot \textbf{C} in \cref{fig:Example_Time_Forecast}). Crucially, this forecast, $\Delta t_{C}$, can be \textit{validated} by when $\mathcal{B}_{500}$ actually makes a detection. In this example, such a detection occurs earlier than expected (plot \textbf{D} in \cref{fig:Example_Time_Forecast}). This inter-window detection error, $\varepsilon$, is used to approximate any second-order effects in the approach towards LOC and is used alongside the individual forecasts of $\mathcal{B}_{300}$ and $\mathcal{B}_{500}$ to produce the final $\Delta t_{LOC}$ estimate as soon as $B_{500}$ makes a detection (plot \textbf{E} in \cref{fig:Example_Time_Forecast}). Finally, $\Delta t_{LOC} = f(\mathcal{S})$ is estimated through a function $f : \mathcal{S} \to \mathbb{R}^{+}$ given a set of individual forecasts, errors, and other parameters. In the example of \cref{fig:Example_Time_Forecast}, $\mathcal{S} = \{ \Delta t_{B}, \Delta t _{D}, \varepsilon\}$. 

The choice of $f(\mathcal{S})$ is flexible and can range from an average to a neural network. In this paper, the inter-window detection error is used to determine a scaling coefficient, $\alpha \in [0, 1]$, to improve estimates of $\Delta t_{LOC}$. Let $\mathcal{D} \in \{D_{1}, D_{2}, ..., D_{d-1}, D_{d}\}$ denote the ordered set of $d$ increasing detection windows of the Bayesian detectors, with $D_{d}$ being the final LOC sensitive window (e.g., $D_{d} = 750$ in the example of \cref{fig:Example_Time_Forecast}). $\alpha$ is then calculated through \cref{eq:forecast_modifier} where $w_{1} \in \mathcal{D}$ and $w_{2} \in \mathcal{D}$ denote the detection windows associated with the inter-window forecasts, subject to $w_{1} < w_{2} < D_{d}$. The true samples at which detections are made are denoted by $i_{w_{1}}$ and $i_{w_{2}}$ respectively. Therefore, a forecast is given by $\Delta\hat{ t} = w_{2} - w_{1}$ whereas the true value is $\Delta t = i_{w_{2}} - i_{w_{1}}$. 
\begin{equation}\label{eq:forecast_modifier}
    \alpha = \text{min}\left ( \frac{\text{max}\left ( i_{w_{2}} - i_{w_{1}}, 0 \right )}{w_{2} - w_{1}}, 1 \right )
\end{equation}

The max$(\cdot)$ in \cref{eq:forecast_modifier} ensures that hindcasting does not occur while the min$(\cdot)$ operator forces $\alpha \leq 1$. The consequence is a conservative forecast that is bounded by the difference between the two largest detection windows, i.e., $\Delta t_{LOC} \in [0, D_{d} - D_{d-1}]$. 

Finally, $\Delta t_{LOC}$ is determined through \cref{eq:delta_t_LOC_applied} where $L = |\mathcal{D} \backslash \{\text{max}(\mathcal{D})\}|$ is the cardinality of $\mathcal{D}$ without $D_{d}$ and $\mathcal{A} \in [0, 1]$ indicates whether a forecast, $\Delta \hat{t}_{LOC}$, is available from a given detector. In the case that the LOC sensitive Bayesian detector with $D_{d}$ makes a detection before any forecasts, then $\Delta t_{LOC} = 0$.  
\begin{equation}\label{eq:delta_t_LOC_applied}
    \Delta t _{LOC} = \frac{\sum_{n}^{L}n\mathcal{A}[n]\alpha[n]\Delta \hat{t}_{LOC}[n]}{\sum_{n}^{L}n\mathcal{A}[n]}
\end{equation}

This completes the \textbf{C}ascaded \textbf{B}ay\textbf{e}sian \textbf{Fore}caster (C-BeFore) forecasting scheme. In practice, we use a collection of early warning signals, $\{E_{1}, ..., E_{K}\}$, to formulate a series of corresponding Bayesian detectors: $\{P(L|E_{1}), ..., P(L|E_{K})\}$. Using these detectors, a $\Delta t_{LOC}$ forecast is obtained through \cref{alg:detection} and \cref{alg:forecast} for the detection and forecasting steps, respectively. In these algorithms, $i \in N$ denotes the current sample with $N \in \mathbb{N}$ the number of data points. The ordered set of increasing detection window sizes is given by $\mathcal{D}$ and contains $d \in \mathbb{N}$ elements. $\lambda \in [0, 1]$ denotes the detection threshold for which detections exceeding this threshold are assigned a value of 1 in $\mathcal{H}$ whereas those below the threshold are assigned 0. $\mathcal{H}$ is thus an $N \times d$ matrix of positive detections. $\mathcal{G}$ denotes an $N \times (d-1)$ matrix of forecast values of each detector, excluding the largest window detector, $\mathcal{B}_{D_{d}}$. Whether a given detector has an active forecast or not is recorded in $\mathcal{A}$, which is a $1 \times d$ array of switches where 0 indicates no forecast available and 1 indicates an available forecast. Even though \cref{alg:detection} is executed first, both algorithms interact with variables which influence each other's behavior between consecutive samples.

At inference, \cref{alg:detection} has a (single core) runtime computational complexity of $O(LKD\text{log}(D))$ where $L$ is the number of detectors, $K$ is the number of $E_{k}$, and $D$ is the largest detection window size. Similarly, \cref{alg:forecast} has a complexity of $O(L + LD)$. Often, $L,K<<D$, thus the complexity of the C-BeFore approach simplifies to $O(D\text{log}(D))$. As a comparison, the (single core) complexity of a recurrent neural network (RNN) operating the same detection window (D) and $E_{k}$ (K) with $H$ hidden neurons is $O(D(KH + H^{2} + H)) \approx O(DH^{2})$. Then, the relative complexity between the C-BeFore and the RNN is $H^{2}/\text{log}(D)$. Therefore, RNNs with $H\geq 3$ neurons are expected to be computationally more complex than the C-BeFore for $D\leq 5000$ samples (which is twice the duration of the yaw LOC inducing maneuver\footnote{In our experience, setting $D$ equal to half the time-to-LOC results in good performance. Thus, $D \approx 500$ (= 1 second) for the quadrotors used here.}).  

Furthermore, we highlight that \cref{alg:detection} and \cref{alg:forecast} transform a series of early detections into forecasts. As such, they can be applied to other early warning signal formulations where it is possible to obtain a series of early detections. This can be used in tandem with some existing fault observers that estimate the loss of effectiveness over moving windows (e.g., \cite{Eltrabyly2022_QuadFTC}) or monitor for growing errors between measured responses and (system model) predictions (e.g, \cite{Fang2020_FDDAutonomousVehicles}).

\begin{algorithm}[!t]
\caption{(Stacked) Bayesian inference loss of control (LOC) \textit{detection algorithm} of the C-BeFore scheme. Here, $|Y|$ denotes the cardinality of a set $Y$.}\label{alg:detection}
\begin{algorithmic}[1]
\Require $\{P(L | E_{1}), ..., P(L | E_{K})\}$, $\mathcal{H}$, $\mathcal{G}$, $\mathcal{D}$, $\mathcal{A}$, $i$, $\lambda$
\State{$\mathcal{H}[i, \text{all}] \leftarrow 0$}
\For {$w$ in range($|\mathcal{D}|)$} \Comment{For each window} 
    \For {$k$ in $K$} \Comment{For each $E_{k}$}
        \State {$p_{w, k} \leftarrow P(L | E_{k, \mathcal{D}[w]})$} \Comment{Detect LOC}
    \EndFor
    \State{$p_{w} \leftarrow$ \cref{eq:stacked_forecasters}} \Comment{Aggregate detections across $E_{k}$} 
    \If{$p_{w} \geq \lambda$} \Comment{Detection exceeds threshold}
        \If{$\mathcal{A}[w] = 0$ \textbf{or} $i > \mathcal{G}[i-1, w]$} \Comment{If forecast is inactive or expired} 
            \State{$\mathcal{H}[i, w] \leftarrow 1$} \Comment{Mark new detection at $i$}
            \State{$\mathcal{A}[w] \leftarrow 0$} 
        \EndIf
    \EndIf
\EndFor
\end{algorithmic}
\end{algorithm}

\begin{algorithm}[!t]
\caption{Time to loss of control (LOC) \textit{forecasting algorithm} of the C-BeFore scheme. Here, $|Y|$ denotes the cardinality of a set $Y$.}\label{alg:forecast}
\begin{algorithmic}[1]
\Require $\mathcal{H}$, $\mathcal{G}$, $\mathcal{D}$, $\mathcal{A}$, $i$
\State{$W \leftarrow \text{argmax}(\mathcal{D})$} \Comment{Index of largest window, $D_{d}$}
\State{$\underline{\mathcal{D}} \leftarrow \mathcal{D} \backslash \{ D_{d}\}$}
\State{$L \leftarrow |\underline{\mathcal{D}}|$}
\If{$\mathcal{H}[i, W] = 1$} \Comment{Largest window detects LOC}
    \State{$\Delta t_{LOC} \leftarrow 0$}
    \State{\textbf{break}}
\Else
    \For {$w$ in range($L$)} \Comment{For each window in $\underline{\mathcal{D}}$}
        \If{$\mathcal{A}[w] = 0$ and $H[i, w] = 1$} \Comment{Check for new detection}
            \State{$\mathcal{G}[i, w] \leftarrow i + (D_{d} - \underline{\mathcal{D}}[w])$} \Comment{Make forecast}
            \State{$\mathcal{A}[w] \leftarrow 1$}
        \Else \Comment{Active forecast exists}
            \State{$\mathcal{G}[i, w] \leftarrow \mathcal{G}[i-1, w]$} \Comment{Hold forecast}
        \EndIf
    \EndFor
    \State{$\boldsymbol{\alpha}[\text{all}] \leftarrow 1$}
    \While{$n < L$} \Comment{Inter-window forecasts}
        \State{$w_{1}, w_{2} \leftarrow \mathcal{D}[n], \mathcal{D}[n+1]$} 
        \State{$i_{w_{1}} \leftarrow \arg\max\limits_{j\in N}\mathcal{H}[\text{all}, n]$}\Comment{True detection index of $w_{1}$}
        \State{$i_{w_{2}} \leftarrow \arg\max\limits_{j\in N}\mathcal{H}[\text{all}, n + 1]$} \Comment{True detection index of $w_{2}$} 
        \State{$\boldsymbol{\alpha}[n] \leftarrow $ \cref{eq:forecast_modifier}} \Comment{Compute modifier}
        \State{$n \leftarrow n + 1$}
    \EndWhile
    \If{sum($\mathcal{H}[i]$) $> \frac{L}{2}$} \Comment{Majority has a forecast}
        \State{$\Delta \hat{t}_{LOC} \leftarrow \mathcal{G}[i, \text{all}] - i$} \Comment{Raw forecasts}
        \State{$\Delta t_{LOC} \leftarrow $ \cref{eq:delta_t_LOC_applied}} \Comment{Aggregate forecasts}
    \EndIf
\EndIf
\end{algorithmic}
\end{algorithm}

\subsection{Data collection \& processing}\label{subsec:data_processing}
\begin{figure}[!t]
    \centering
    \subfloat[CineGo]{\label{fig:CineGo}\includegraphics[width=.49\linewidth]{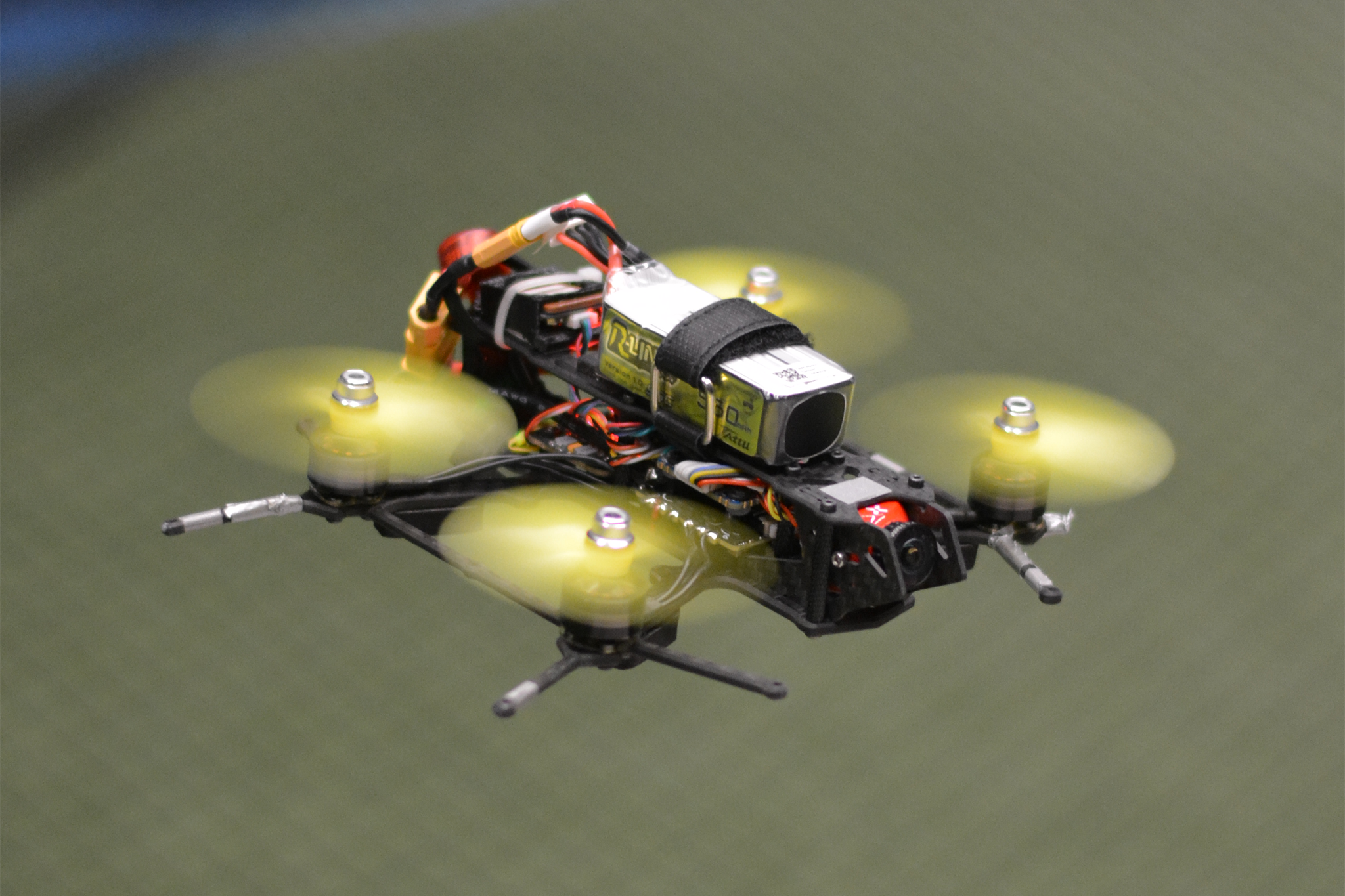}}\hfill%
    \subfloat[DataCan75]{\label{fig:DataCan75}\includegraphics[width=.49\linewidth]{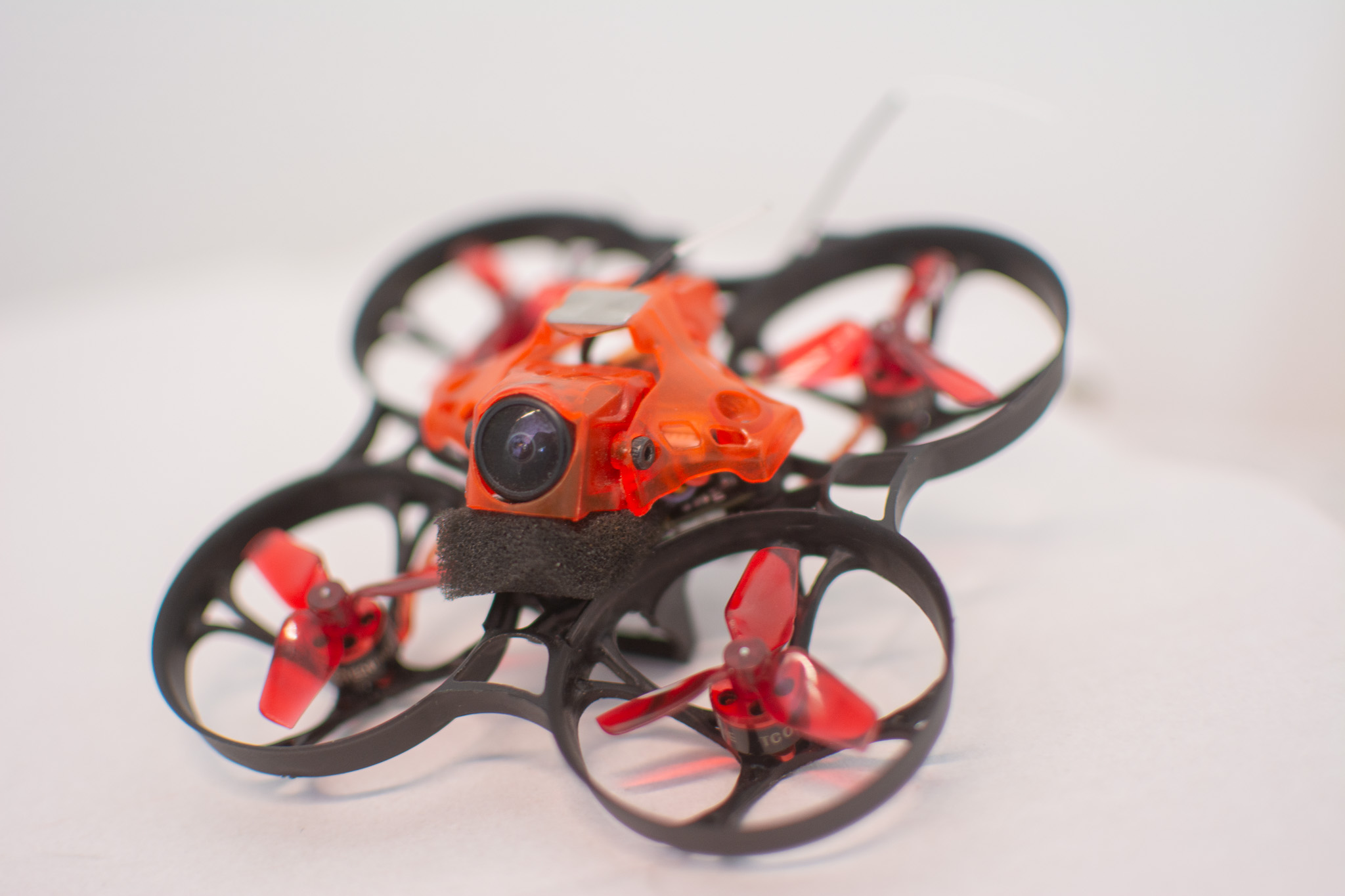}}
    \caption{Quadrotor platforms used to collect yaw loss of control data \cite{Altena_ANN_LOC}.}
    \label{fig:quadrotors}
\end{figure}

The yaw-induced LOC data is collected using two quadrotor platforms, the CineGo (\cref{fig:CineGo}) and the DataCan75 TinyWhoop (\cref{fig:DataCan75}), built from off-the-shelf components (see \cref{tab:quad_properties}). Both quadrotors run the Betaflight 4.2 flight control software, which was chosen in \cite{Altena_ANN_LOC} as it accommodates high frequency logging of the rotor speed (e)RPM measurements through Bidirectional DShot. Flight data is logged at 500Hz on both quadrotors.

\begin{table*}[!t]
\caption{Properties of the quadrotor platforms.}
\label{tab:quad_properties}
\centering
\scriptsize
\begin{tabular}{@{}p{3.8cm}p{2.9cm}p{2.4cm}p{2.9cm}p{2.9cm}@{}} 
\toprule
& \textbf{CineGo} (\cref{fig:CineGo}) & \textbf{DataCan75} (\cref{fig:DataCan75}) & \textbf{SuperKnight5} (\cref{fig:SK5}) & \textbf{DamselFly} (\cref{fig:DamselFly}) \\ \midrule
\textbf{Mass (incl. batteries), $g$} & 265 & 56 & 480 & 510\\
\textbf{Diagonal hub-to-hub diameter, $mm$} & 155 & 75 & 200 & 153 \\
\textbf{Propeller diameter, $mm$} & 76 & 35 & 127 & 76\\
\textbf{Motor} & Emax Eco 1407 3300 kV & TC0803 15000 kV & T-motor F40 Pro IV 1950 kV & Emax Eco 1407 4100 kV \\
\textbf{Batteries} & Tattu R-Line 14.8V 550mAh & 2x BETAFPV 3.45V 300mAh & Tattu FunFly 22.2V 1300mAh & Tattu FunFly 22.2V 1300mAh\\
\textbf{Flight Controller (FC)} & MATEKSYS F722-mini 2-8S & JHEMCU SH50 F4 2S & iFlight Beast H743-AIO &  	MATEKSYS H743-SLIM V3 \\
\textbf{FC Software} & Betaflight 4.2 & Betaflight 4.2 & Betaflight 4.3 & Indiflight\\
\bottomrule
\end{tabular}
\end{table*}

In the yaw LOC data set of \cite{Altena_ANN_LOC}, not all flights with the yaw LOC maneuver meet the attitude definition of LOC (i.e., \cref{eq:LM-ATT}) before a crash occurs, which can happen due to the confines of the flight arena. We remove such flights from the nominal (i.e., non-lOC) data set in the subsequent analysis for consistency. Instead, the nominal data set contains flights which include aggressive (up to 50 m/s$^{2}$ accelerations and 500 deg/s pitch \& roll rotational rates) and high-speed maneuvers (up to 8 m/s).

\begin{figure}[!t]
    \centering
    \includegraphics[width=0.8\linewidth]{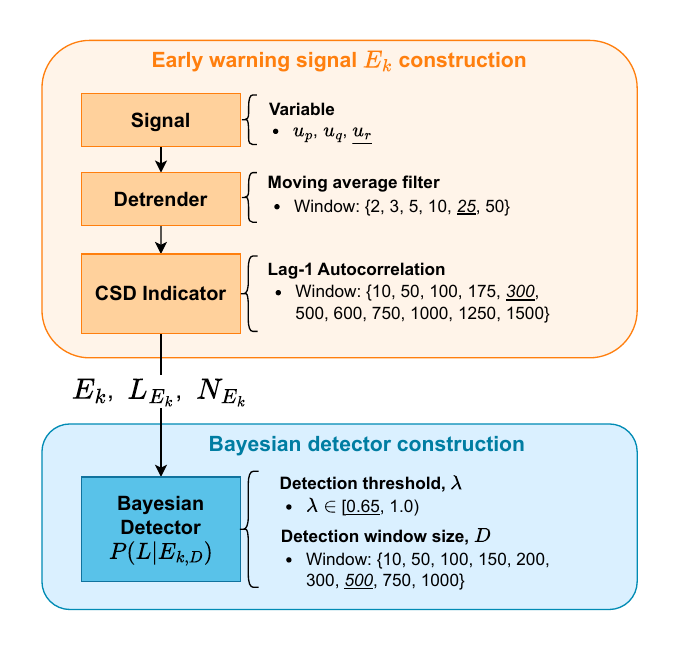}
    \caption{Parameters of the early warning signal, $E_{k}$, and Bayesian loss of control detector, $P(L | E_{k, D})$. An example configuration is \underline{underlined}.}
    \label{fig:EWS_BID_Design}
\end{figure}

The raw flight data is resampled to 500 Hz in order to account for inconsistencies in the logging rate. The rotor speed difference measurements (i.e., \cref{eq:rotor_diff}) are then detrended prior to computing the critical slowing down AC1 indicator. Subsequently, the LOC ($L_{E_{k}}$) and nominal ($N_{E_{k}}$) distributions are created as described in \cref{subsec:EWS_design}. An overview of the parameters involved with the $E_{k}$ and Bayesian detector design can be found in \cref{fig:EWS_BID_Design}. A parame ter sweep analysis is used to select suitable parameters, shown in \cref{tab:Ek_and_Detector_Params}. All processing steps are conducted in Python.

\setlength\tabcolsep{3pt}
\begin{table}[!t]
\caption{Early warning signal, $E_{k}$, and Bayesian detector, $P(L | E_{k, D})$, parameters.}
\label{tab:Ek_and_Detector_Params}
\centering
\scriptsize
\begin{tabular}{lcc} 
\toprule
\textbf{Quadrotor} & DataCan75 & CineGo \\ \midrule
$\mathbf{E_{k}}$ \textbf{parameters} \\
\begin{tabular}{@{}l@{}}Signal(s)\end{tabular} & 
$\{ u_p, u_q, u_r\}$ &  $\{ u_p, u_q, u_r\}$ \\[0.5ex]
\begin{tabular}{@{}l@{}}Moving average window (samples)\end{tabular} & 
10 & 25 \\[0.5ex]
\begin{tabular}{@{}l@{}}Autocorrelation window (samples) \end{tabular} & 50 & 50 \\ \midrule
$\mathbf{P(L | E_{k, D})}$ \textbf{parameters} \\
\begin{tabular}{@{}l@{}}Detector window sizes (samples) \end{tabular} &  [200, 500, 750] &  [150, 500, 750] \\[0.5ex]
\begin{tabular}{@{}l@{}}Forecast threshold $\lambda$\end{tabular} & 
0.89 & 0.88 \\
\bottomrule
\end{tabular}
\end{table}

\section{RESULTS AND DISCUSSION}\label{sec:results}
\subsection{Baseline comparison}\label{subsec:RNNs_vs_C-BeFore}

As a baseline comparison, the C-BeFore developed in this work are compared to the recurrent neural network (RNNs) of \cite{Altena_ANN_LOC}. This baseline comparison is limited to the DataCan75 quadrotor as these RNNs were trained using only this data set \cite{Altena_ANN_LOC}. The DataCan75 data set consists of 49 loss of control (LOC) flights supplemented by 6 nominal (i.e., non-LOC) flights. Moreover, in \cite{Altena_ANN_LOC}, each RNN architecture (LSTM, BiLSTM, CNNLSTM, and GRU) was trained over multiple initializations to mitigate training biases. Of these, the training runs with the best LOC classification performance are compared to the C-BeFore.

\begin{figure}[!t]
    \centering
    \includegraphics[width = \linewidth]{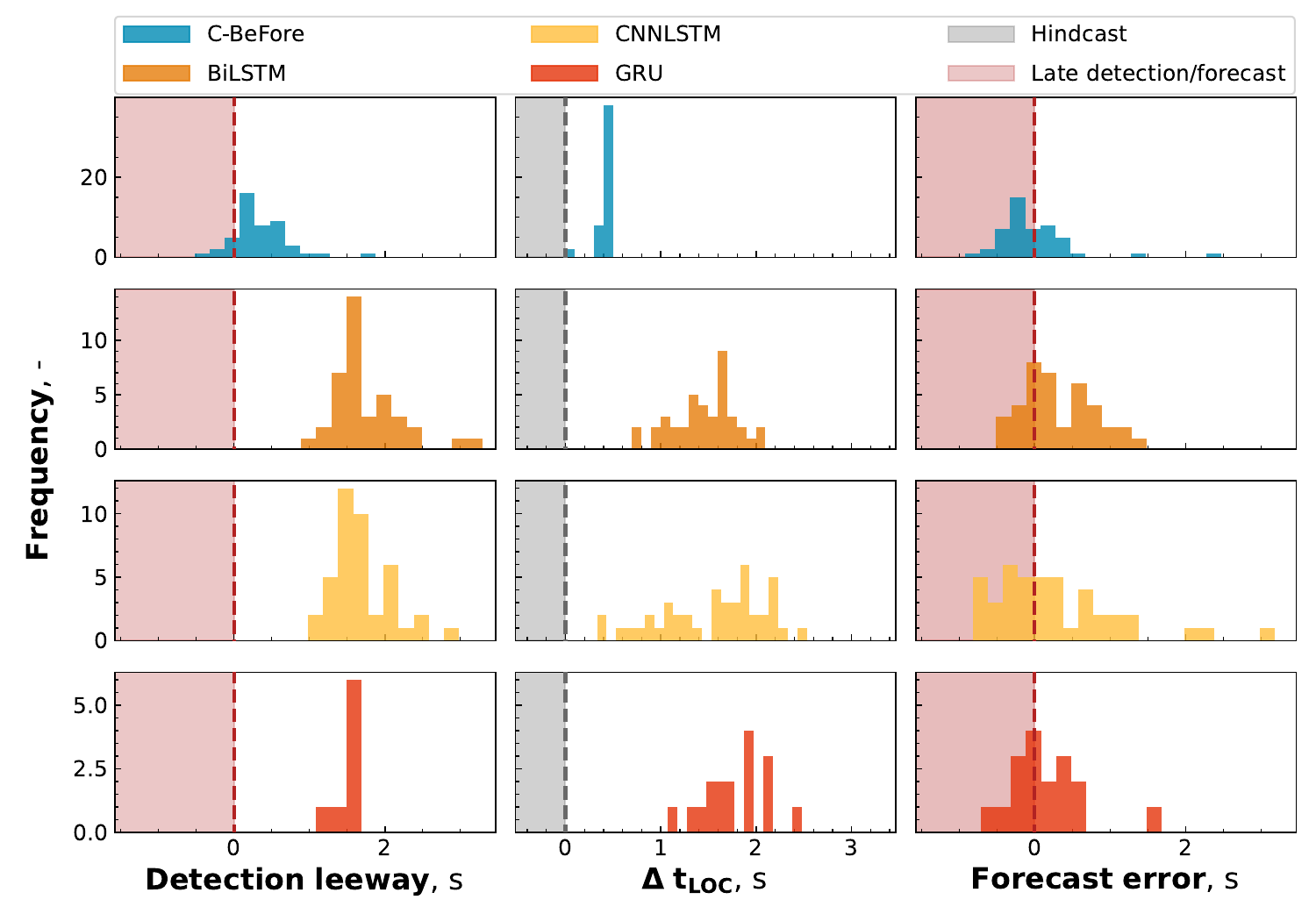}
    \caption{Loss of control (LOC) forecasting performance comparison between the DataCan75 C-BeFore of \cref{tab:Ek_and_Detector_Params} and the BiLSTM, CNNLSTM, and GRU neural networks of \cite{Altena_ANN_LOC}. The plots in the left column depict how early a detection is with respect to the labeled LOC moment. The associated time-to-LOC forecasts, $\Delta t_{LOC}$, are illustrated by the plots in the middle column. The right column plots show the forecast errors.} 
    \label{fig:DataCan75_ForecastPerf_Comp}
\end{figure}

\Cref{tab:Forecasting_comparison,fig:DataCan75_ForecastPerf_Comp} compare the performance of the RNNs and C-BeFore. While none of the RNNs LOC detections are late, the subsequent $\Delta t_{LOC}$ forecasts are not reliable: they fail to consistently distinguish between LOC and nominal quadrotor behavior. Moreover, the RNNs have a tendency to produce more early false positives - that is, a detection of LOC before the dangerous yaw maneuver is even initiated - than the C-BeFore, as shown in \cref{tab:Forecasting_comparison}. The C-BeFore reduces this false positive rate by at least 83\%, at the cost of a shorter $\Delta t_{LOC}$. Even so, the average $\Delta t_{LOC} \approx 0.44$ seconds afforded by the C-BeFore is sufficient to perform corrective actions and prevent LOC\footnote{\label{fn:1}The DataCan75 Betaflight control loop runs at 4kHz.}.

\begin{sidewaystable}
\setlength\tabcolsep{3pt}
\caption{Loss of control (LOC) classification and forecasting accuracies for the quadrotors in \cref{tab:quad_properties}.}
\label{tab:Forecasting_comparison}
\centering
{\setlength{\extrarowheight}{3pt}
\begin{tabular}{lccccccc} 
 \begin{tabular}{l}\textbf{Quadrotor} (\textit{flight controller}) \\ LOC forecaster \end{tabular}  &\begin{tabular}{c}No. True \\ positives\end{tabular} & \begin{tabular}{c}No. False \\ negatives\end{tabular} & \begin{tabular}{c}No. True \\ negatives\end{tabular} & \begin{tabular}{c}No. (SysID.) \\ False positives\end{tabular} & \begin{tabular}{c}No. (Early) \\ False positives\end{tabular} & \begin{tabular}{c}No. Instant \\ forecasts \end{tabular} & \begin{tabular}{c}Forecast error \\ \textit{Mean (Std.)}\end{tabular}\\[8pt] \hline \hline
\textbf{DataCan75} (\textit{Betaflight}) & & & & & & & \\ \hline
BiLSTM [Run 31]  & 40 & 0 & 4 & 2 & 9 & 0 & 0.388 (0.712)\\
LSTM [Run 14]  & 41 & 0 & 0 & 6 & 8 & 0 & 0.353 (0.649)\\
CNNLSTM [Run 31]  & 43 & 0 & 2 & 4 & 6 & 0 & 0.252 (0.831)\\
GRU [Run 30]  & 17 & 0 & 4 & 2 & 32 & 0 & 0.133 (0.4967)\\ 
Data-driven C-BeFore & 48 & 0 & 5 & 1 & 1 & 2 & -0.027 (0.515) \\ \hline
\multicolumn{1}{r}{Total:} & $\cdot / 49$ & $\cdot / 6$ & $\cdot / 6$ & $\cdot / 6$ & $\cdot / 49$ & $\cdot / 49$ &
\\
\textbf{CineGo} (\textit{Betaflight}) & & & & & & & \\ \hline
Data-driven C-BeFore   & 34 & 0 & 9 & 2 & 0 & 0 & 0.199 (0.247)\\
Assumed C-BeFore   & 34 & 0 & 9 & 2 & 0 & 0 & 0.172 (0.267)\\ \hline
\multicolumn{1}{r}{Total:}& $\cdot / 34$ & $\cdot / 11$ & $\cdot / 11$ & $\cdot / 11$ & $\cdot / 34$ & $\cdot / 34$  &
\\
\textbf{SuperKnight5} (\textit{Betaflight}) & & & & & & & \\ \hline
Assumed C-BeFore$^{*}$  & 4 & 0 & 0 & 1 & - & 0 & -3.139 (1.671)\\ \hline
\multicolumn{1}{r}{Total:} & $\cdot / 4$ & $\cdot / 1$ & $\cdot / 1$ & $\cdot / 1$ & $\cdot / 4$ & $\cdot / 4$ & 
\\
\textbf{DamselFly} (\textit{Indiflight})& & & & & & &  \\ \hline
Assumed C-BeFore$^{*}$  & 4 & 0 & 12 & 1 & - & 0 & -0.718 (0.039)\\ \hline
\multicolumn{1}{r}{Total:} & $\cdot / 4$ & $\cdot / 13$ & $\cdot / 13$ & $\cdot / 13$ & $\cdot / 4$ & $\cdot / 4$ & \\
\hline \hline
\end{tabular}}
\begin{flushleft}
$^{*}$Assumed C-BeFore parameterized for the CineGo quadrotor and LOC data, applied without additional modification or re-parameterization. 
\end{flushleft}
\end{sidewaystable}

\subsection{Forecasting under data-informed distributions}\label{subsec:fitted_LOC_forecast}
The CineGo LOC data set, consisting of 34 yaw LOC and 11 non-LOC flights, is used to compare the performance of the C-BeFore when the early warning signal behavior near LOC is known (data-driven) versus when no data is available and only assumptions on its behavior can be made (assumed). This data set is chosen as more nominal flights are available for comparison. Here, we present the results for the data-driven C-BeFore whereas the results for the assumed variant are given in \cref{subsec:assumed_LOC_forecast}.

\begin{figure}[!t]
    \centering
    \includegraphics[width = \linewidth]{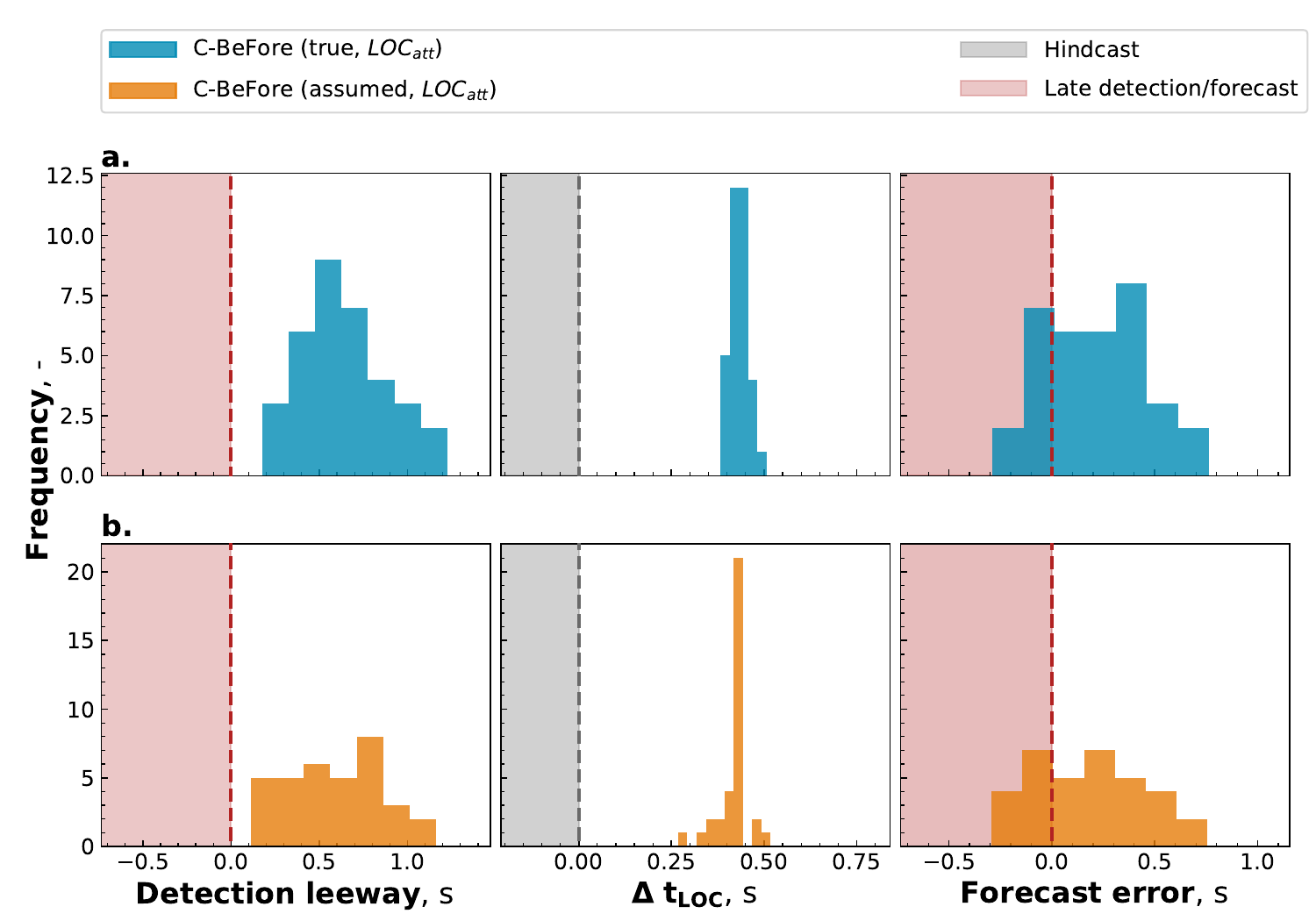}
    \caption{Loss of control (LOC) forecasting performance comparison between the data-driven and assumed C-BeFore for the CineGo yaw-induced LOC data set. The left column plots indicate the time between a detection of LOC and its label. The associated time-to-LOC forecasts, $\Delta t_{LOC}$, is shown in the middle column plots. The right column plots illustrate the associated forecast errors.}
    \label{fig:CineGo_ForecastPerf}
\end{figure}

The forecasting performance of the LOC data-driven C-BeFore is shown in \textbf{a} of \cref{fig:CineGo_ForecastPerf}, indicating that all LOC events are detected before they occur. This is implies that the 90 degree attitude condition of $LOC_{att}$ has not yet been reached and may still be avoided. On average, the LOC detections occur before roll ($\phi$) or pitch ($\theta$) attitude exceed a magnitude of 40 degrees with most (i.e., 75\%) of these occurring below 50 degrees. Though these attitude angles are arguably high, they are common during aggressive maneuvering: $\phi$ and $\theta$ angles exceeding 70 degrees were routinely reached during the nominal flights of the CineGo. Crucially, the majority of the LOC forecasts occur well before the critical 90 degree attitude condition and remain well within the flight envelope of the quadrotor, which improves chances of recovery. 

Furthermore, the LOC data-driven C-BeFore achieves a 100\% true positive rate (34/34 LOC flights) with zero early false positives (see \cref{tab:Forecasting_comparison}). Most (82\%) of the nominal flights are correctly recognized as non-LOC.

\subsection{Forecasting under assumed distributions}\label{subsec:assumed_LOC_forecast}

\begin{figure}[!t]
    \centering
    \includegraphics[width=\linewidth]{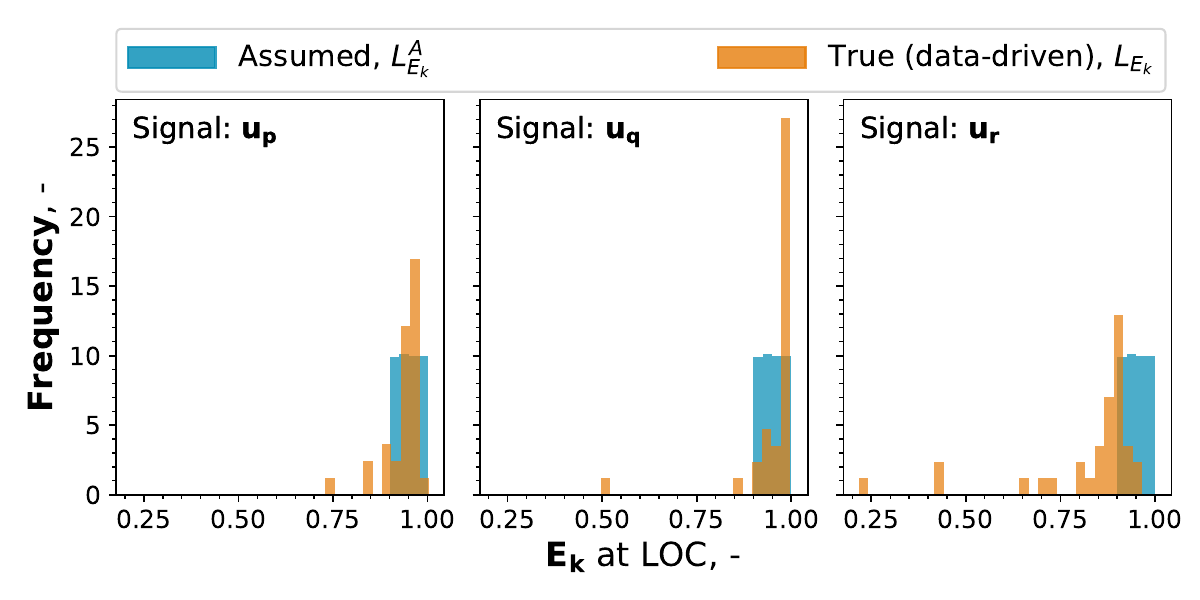}
    \caption{Comparison between the assumed, $L^{A}_{E_{k}}$, and the true, $L_{E_{k}}$, loss of control (LOC) distributions of the CineGo yaw-induced LOC data, for each early warning signal ($E_{k}$): $u_{p}$, $u_{q}$, and $u_{r}$.}
    \label{fig:TrueVsAssumed_LOC_distribution_comparison}
\end{figure}

Here, we show that the C-BeFore can still provide practical LOC forecasts without data of the LOC event. We substitute the true LOC distributions of the CineGo C-BeFore with an assumed LOC distribution, $L_{E_{k}}^{A}$. As an extreme example, we take $L_{E_{k}}^{A}$ to be a constant distribution on the interval $[0.9, 1.0]$ in $E_{k}$ (see \cref{fig:TrueVsAssumed_LOC_distribution_comparison} for a comparison with the true LOC distribution). This choice of interval is motivated by the expected behavior of $E_{k}$ as LOC is approached following critical slowing down theory.

The performance of the assumed C-BeFore is presented in \cref{tab:Forecasting_comparison} and \cref{fig:CineGo_ForecastPerf}. Overall, there is almost no difference in performance between the assumed and LOC data-driven C-BeFore. The cost of using the assumed distribution is that the detections and forecasts occur (slightly) later. Even so, the forecasts remain useful and suggest that our approach does not depend on extensive LOC data sets.

\subsection{Generalization capacity}\label{subsec:generic_capacity}

In this section, we demonstrate that the C-BeFore generalizes well to other quadrotors and LOC scenarios. For this, we choose the CineGo parameterized assumed C-BeFore that relies on the assumed distribution ($L_{E_{k}}^{A}$) of LOC behavior. We apply this forecaster \textit{without any re-parameterization} to now detect flyway events that occur on other quadrotors - the SuperKnight5 and DamselFly in \cref{tab:quad_properties} - flying both outdoors and indoors in windy conditions. Moreover, both these quadrotors employ different controllers from the CineGo.

\subsubsection{Flyaway flight data}
\begin{figure}[!t]
    \centering
    \subfloat[SuperKnight5]{\label{fig:SK5}\includegraphics[width=.49\linewidth]{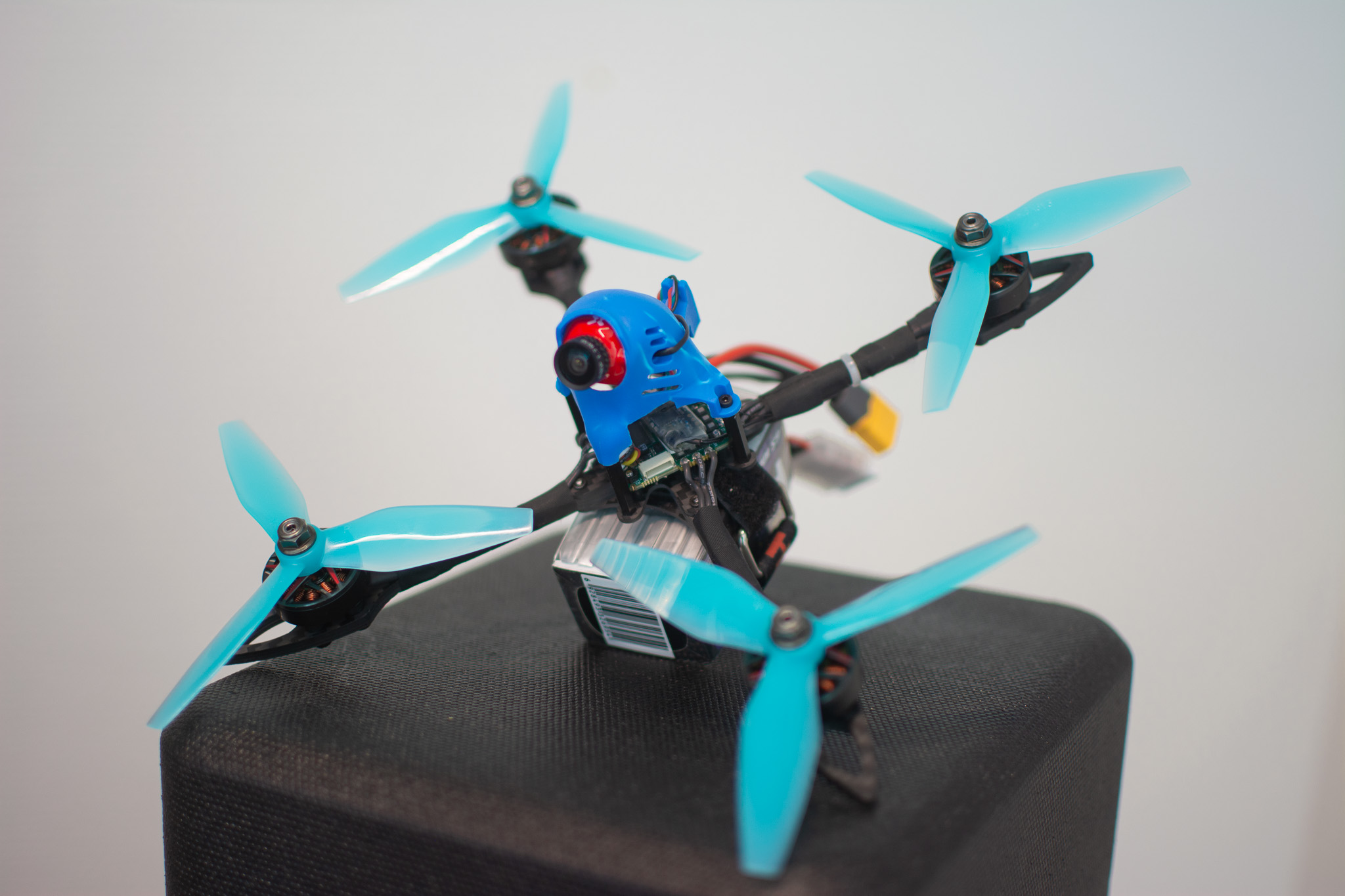}}\hfill%
    \subfloat[DamselFly]{\label{fig:DamselFly}\includegraphics[width=.49\linewidth]{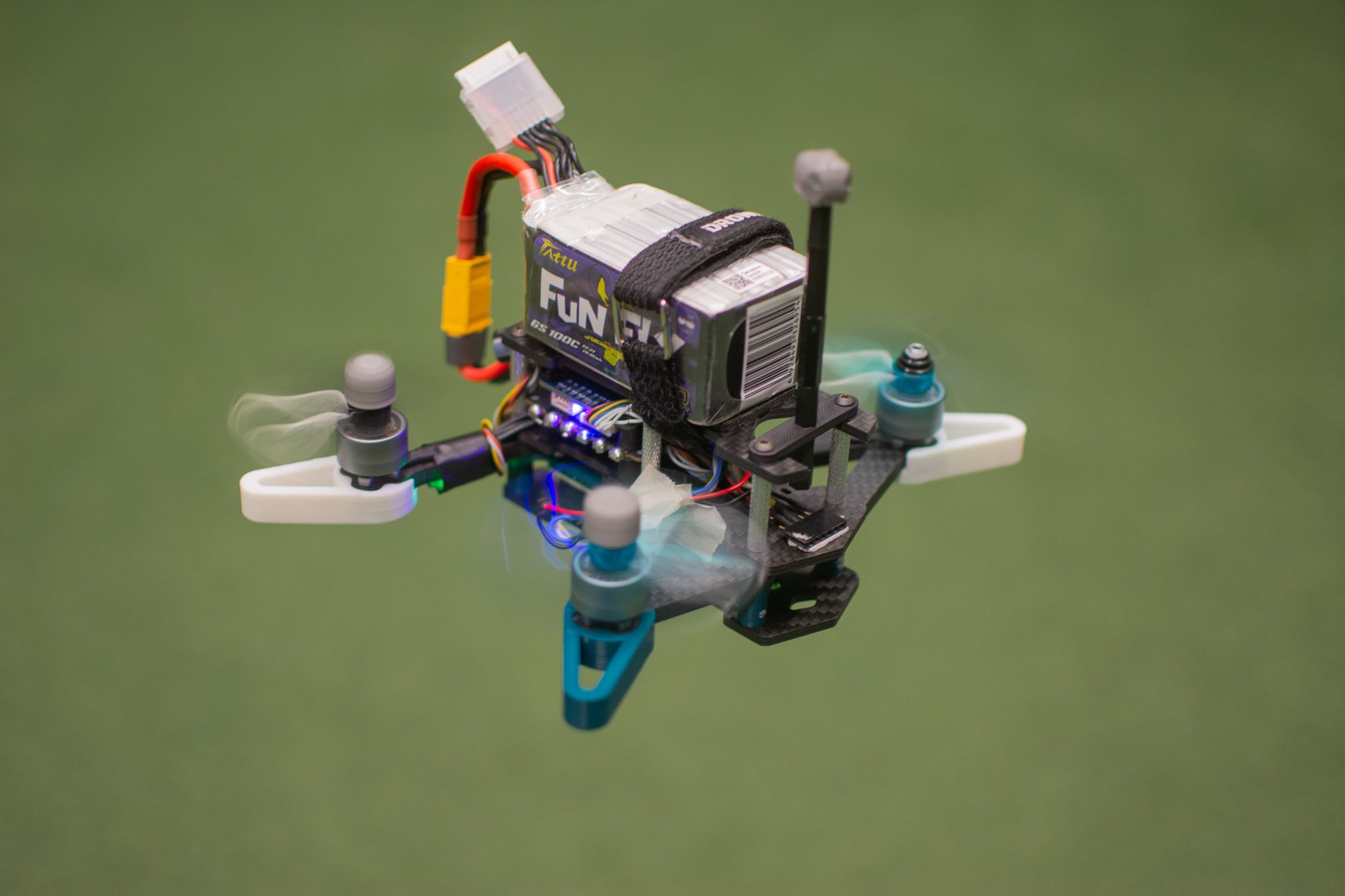}}
    \caption{Quadrotors that experience flyaways.}
    \label{fig:quadrotors_validation}
\end{figure}

The SuperKnight5 (\cref{fig:SK5}) was designed as a high thrust-to-weight ratio ($>$11) outdoor FPV quadrotor. However, it suffered from structural resonance that in turn caused excited controller instabilities, resulting in flyaways (uncontrollable ascending flight) at certain regions of its operational envelope (see \cref{fig:flyaway_sk5}). These flyaways occurred on 4 out of the 5 outdoor flights during the flight testing campaign over several days, design iterations, and controller tuning. Each flight was manually piloted (using BetaFlight's angular rate mode\footnote{Meaning that the pilot commands the body rates of the quadrotor - $p$, $q$, and $r$ - alongside the total thrust.}) and several aggressive maneuvers were performed (e.g., 11g punch-out). The SuperKnight5 uses Betaflight 4.3 and logs the rotor speed measurements at 500 Hz. Thus, the exact same CineGo $E_{k}$ data processing steps (\cref{subsec:data_processing}) and parameters in \cref{tab:Ek_and_Detector_Params} are reused here. 

Conversely, the DamselFly (\cref{fig:DamselFly}) operates the Indiflight\footnote{Available here: \url{https://github.com/tudelft/Indiflight}} flight controller, which differs from Betaflight in that the inner-loop control architecture uses incremental non-linear dynamic inversion (INDI) as opposed to a series of (adaptive) PIDs, allowing for autonomous flight. As INDI is a sensor-based approach, excessive noise in the measurements (e.g., arising from rotor vibrations) can induce flyaways when coupled with (sudden) disturbances, such as wind gusts. This occurred in 4 of the 17 indoor autonomous flights of the DamselFly. These flights involved position and trajectory tracking tasks (at speeds of up to 5 m/s) subject to wind disturbances produced by a large fan. In Indiflight, the rotor speed measurements are recorded at 1000 Hz but are resampled to 500 Hz to follow the same data processing steps as in \cref{subsec:data_processing}. Moreover, since Indiflight uses a different filter setup than Betaflight, the $E_{k}$ for the DamselFly are computed using different detrending and lag-1 autocorrelation (AC1) windows to ensure adequate detrending and $E_{k}$ sensitivity: a moving average window of 8 samples and an AC1 window of 25 samples are used.

Additionally, for both the DamselFly and the SuperKnight5, the nominal distribution $N_{E_{k}}$ is derived from their nominal flight data (and not the CineGo) as it is reasonable to assume that such nominal data is available. All other forecaster parameters (i.e., forecasting windows, detection thresholds, and $L_{E_{k}}^{A}$ distributions) are kept as-is from the CineGo.

\subsubsection{Flyaway detections}

The flyaway classification performance of the CineGo C-BeFore applied to both the SuperKnight5 and DamselFly data sets is shown in \cref{tab:Forecasting_comparison}. Despite the differences in controllers and quadrotor properties, the CineGo C-BeFore successfully detects \textit{all} flyaway events that occur on the SuperKnight5 and DamselFly, even though there are no significant off-axis (i.e., roll or pitch) oscillations characteristic of the yaw LOC scenario used to parameterize the forecaster. These results suggest that the proposed approach is indeed sensitive to the evolving controller-induced instabilities common in both the flyaways and the yaw LOC scenario. Furthermore, the C-BeFore correctly classifies most (92\%) of the nominal flights as non-LOC for the DamselFly, despite the difference in controller architecture (i.e., autonomous Indiflight vs human-piloted Betaflight).

\begin{figure}[!t]
    \centering
    \includegraphics[width=\linewidth]{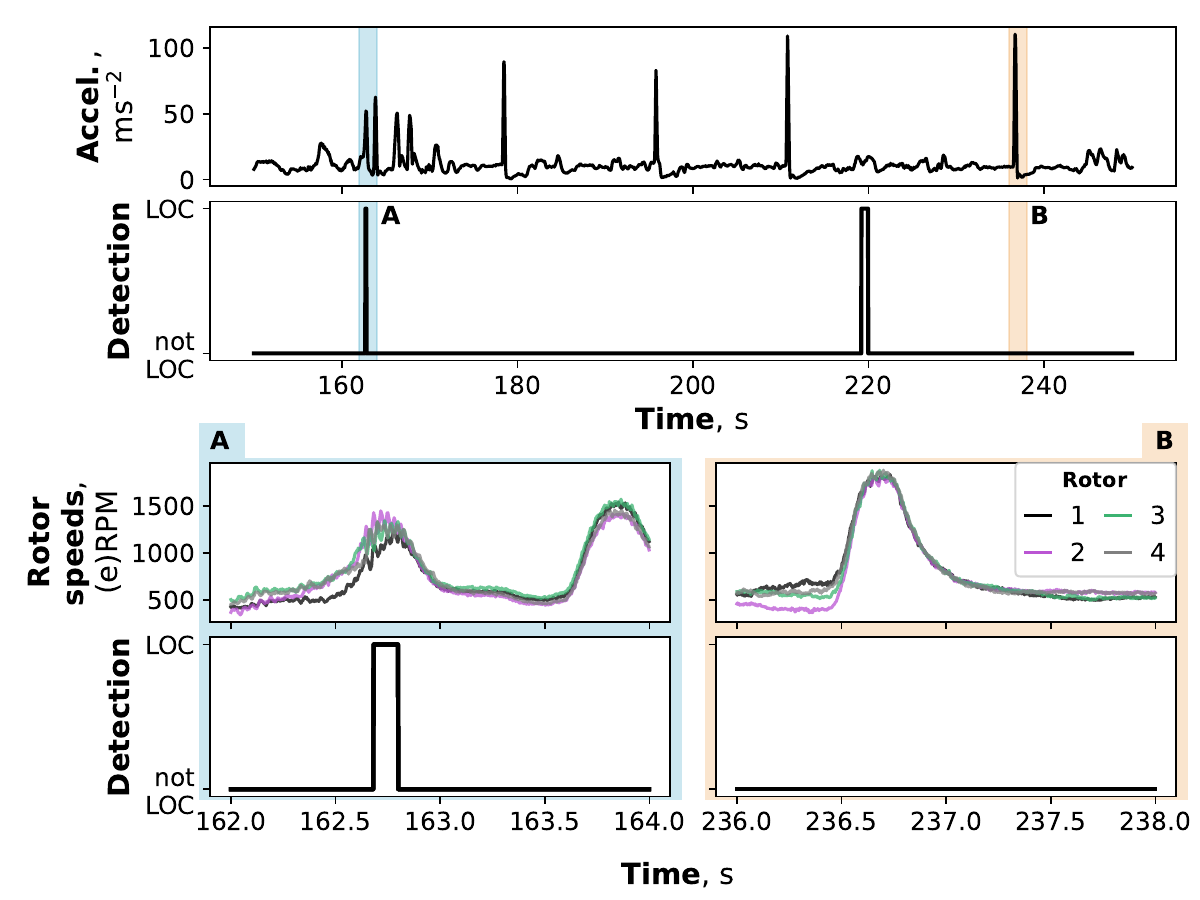}
    \caption{Loss of control (LOC) detections CineGo C-BeFore applied on the `nominal' outdoor flight data of the SuperKnight5. The blue shaded region (A) highlights a false positive detection due to oscillations in the rotor speed response. The orange shaded region (B) shows one of the 11g ($\approx 108 ms^{-2}$) punch out maneuvers.}
    \label{fig:SK5_example_detection}
\end{figure}

\begin{figure}[!t]
    \centering
    \includegraphics[width=\linewidth]{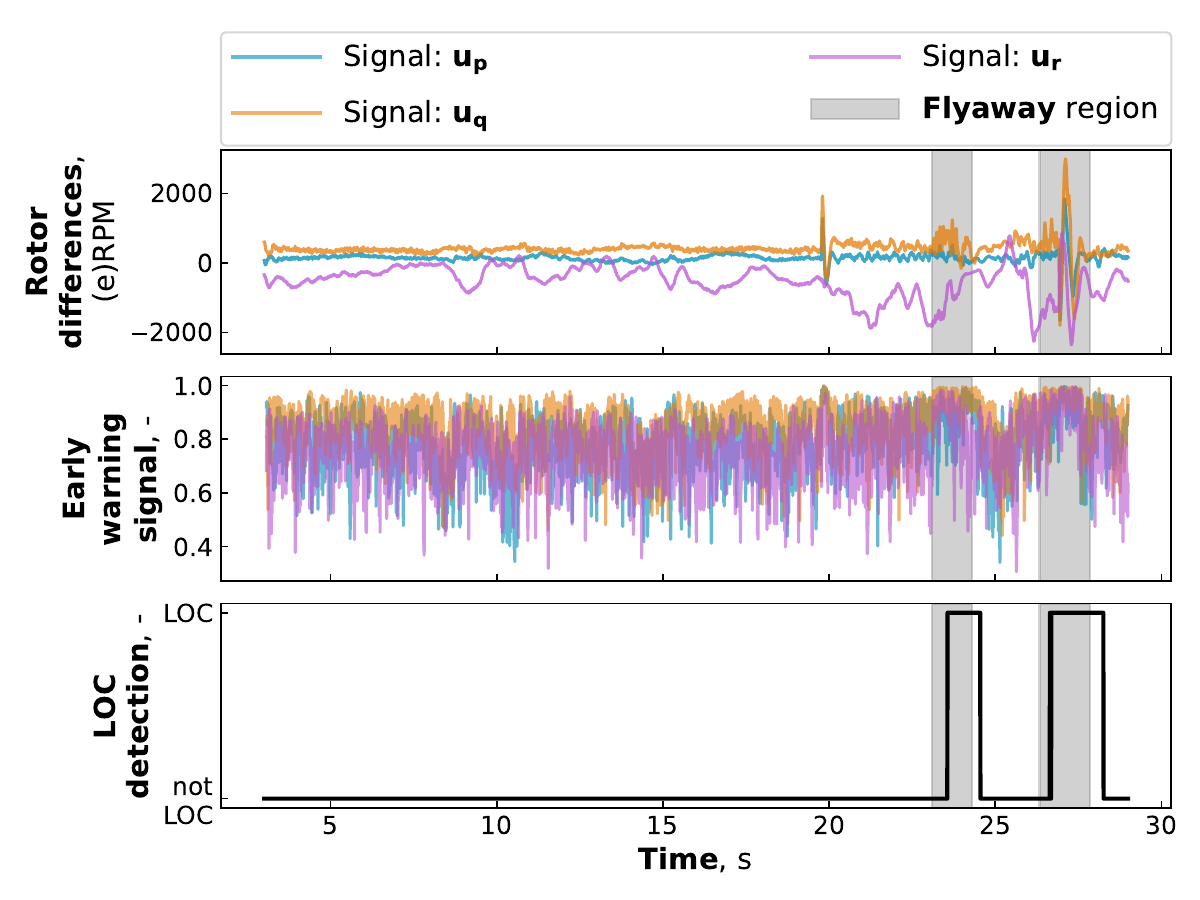}
    \caption{Consecutive flyaway detections made by the CineGo parameterized C-BeFore as applied to real flight data of the DamselFly quadrotor.}
    \label{fig:Dfly_Flyaway_detection}
\end{figure}

While there are false positive detections that arise for the SuperKnight5, these do not occur during any of the aggressive 11g punch out maneuvers. Instead, the detections follow from episodes of concerning oscillations in rotor speeds (compare \textbf{A} with \textbf{B} in \cref{fig:SK5_example_detection}). These oscillations are undesirable and often result in flyaways which are promptly detected by the C-BeFore. This ability is apparent also for the DamselFly, where consecutive flyaways are promptly detected in \cref{fig:Dfly_Flyaway_detection}.

\subsection{Limitations}

While the proposed forecasting approach does not explicitly require loss of control (LOC) data, any knowledge of the LOC event works to inform the early warning signal (EWS) and Bayesian detector design. Likewise, in order to obtain practical forecasts of LOC, the time-scale of the LOC event is needed to properly constrain the detector observation windows.

Furthermore, a key challenge with applying the proposed approach to other systems lies in the development of suitable EWS. Obtaining adequate detrending remains a standing issue in the CSD literature. When no LOC data is available, we advise that the EWS be designed such that the nominal distribution is closely centered around 0.5 in the lag-1 autocorrelation metric which - in our experience - seems to promote tipping point sensitivity. Moreover, in this paper we rely on rotor speed measurements to construct the EWS, which can limit compatible autopilots. However, recent updates to popular autopilots, such as Px4\footnote{See \url{https://docs.px4.io/main/en/peripherals/dshot.html}.}, are introducing such support. 

Finally, while our approach works to anticipate LOC in a generic manner, it does not provide corrective control actions to help prevent LOC nor does it identify the source that drives the approach towards LOC. This is the trade-off made by generic nature of the approach: no system model or LOC data is needed to detect LOC, and thus, there is little system-specific information available to help diagnose the issue. Nonetheless, the the ability to identify that a LOC event is approaching is an important first step towards its prevention. Future work should help derive preventative measures while retaining the generic nature of the approach. 

\section{CONCLUSION}\label{sec:conclusion}
In this paper, we propose a forecasting framework to anticipate controller-induced quadrotor loss of control (LOC) events. We apply this forecaster to real quadrotor flight data wherein LOC arises from unstable controller behavior. The proposed approach outperforms state-of-the-art recurrent neural network (RNN) quadrotor LOC forecasters in terms of classification accuracy between LOC and aggressive, yet nominal, flights. Moreover, the approach is LOC data-free as it does not involve a training phase nor does it require a LOC data set to produce informative forecasts and detections. This sensitivity for LOC events is also shown to generalize to other controllers and quadrotor platforms.  

Future work will apply the the approach to anticipate critical transitions and LOC occuring in other (autonomous) systems. To facilitate this, ways that simplify the EWS design process and the selection of suitable forecasting parameters for the system of interest should be explored. While improving the forecasting performance and usability of our approach is desirable, the ultimate goal is to prevent LOC entirely. Thus, future work will also apply the approach for the purposes of proactive LOC prevention.

\section*{Funding Sources}

This work is supported through The Dutch Research Council (NWO) VIDI grant 18378 on ``forecasting safe operating envelopes for autonomous robots". 

\bibliography{sample}

@INPROCEEDINGS{Sun2020_UpsetRecovery,
    author={S. {Sun} and M. {Baert} and B. S. {van Schijndel} and C. C. {de Visser}},
    booktitle={2020 IEEE International Conference on Robotics and Automation (ICRA)}, 
    title={Upset Recovery Control for Quadrotors Subjected to a Complete Rotor Failure from Large Initial Disturbances}, 
    year={2020},
    volume={},
    number={},
    pages={4273-4279},
    doi={10.1109/ICRA40945.2020.9197239},
    ISSN={2577-087X},
    month={5},
}

@article{Scheffer2009,
   author = {Marten Scheffer and Jordi Bascompte and William A. Brock and Victor Brovkin and Stephen R. Carpenter and Vasilis Dakos and Hermann Held and Egbert H. Van Nes and Max Rietkerk and George Sugihara},
   doi = {10.1038/nature08227},
   issn = {1476-4687},
   issue = {7260},
   journal = {Nature 2009 461:7260},
   month = {9},
   pages = {53-59},
   pmid = {19727193},
   publisher = {Nature Publishing Group},
   title = {Early-warning signals for critical transitions},
   volume = {461},
   url = {https://www.nature.com/articles/nature08227},
   year = {2009},
}

@article{Brett2020,
   author = {Tobias Brett and Marco Ajelli and Quan Hui Liu and Mary G. Krauland and John J. Grefenstette and Willem G. Van Panhuis and Alessandro Vespignani and John M. Drake and Pejman Rohani},
   doi = {10.1371/JOURNAL.PCBI.1007679},
   isbn = {1111111111},
   issn = {1553-7358},
   issue = {3},
   journal = {PLOS Computational Biology},
   pages = {e1007679},
   pmid = {32150536},
   publisher = {Public Library of Science},
   title = {Detecting critical slowing down in high-dimensional epidemiological systems},
   volume = {16},
   url = {https://journals.plos.org/ploscompbiol/article?id=10.1371/journal.pcbi.1007679},
   year = {2020},
}

@article{Maturana2020,
   author = {Matias I. Maturana and Christian Meisel and Katrina Dell and Philippa J. Karoly and Wendyl D’Souza and David B. Grayden and Anthony N. Burkitt and Premysl Jiruska and Jan Kudlacek and Jaroslav Hlinka and Mark J. Cook and Levin Kuhlmann and Dean R. Freestone},
   doi = {10.1038/s41467-020-15908-3},
   issn = {2041-1723},
   issue = {1},
   journal = {Nature Communications 2020 11:1},
   month = {5},
   pages = {1-12},
   pmid = {32358560},
   publisher = {Nature Publishing Group},
   title = {Critical slowing down as a biomarker for seizure susceptibility},
   volume = {11},
   url = {https://www.nature.com/articles/s41467-020-15908-3},
   year = {2020},
}

@inproceedings{belcastro2012loss,
  title={Loss of control prevention and recovery: Onboard guidance, control, and systems technologies},
  author={Belcastro, Christine},
  booktitle={AIAA Guidance, Navigation, and Control Conference},
  pages={4762},
  year={2012}
}

@article{Dakos2012,
   author = {Vasilis Dakos and Egbert H. Van Nes and Paolo D'Odorico and Marten Scheffer},
   doi = {10.1890/11-0889.1},
   issn = {1939-9170},
   issue = {2},
   journal = {Ecology},
   month = {2},
   pages = {264-271},
   pmid = {22624308},
   publisher = {John Wiley & Sons, Ltd},
   title = {Robustness of variance and autocorrelation as indicators of critical slowing down},
   volume = {93},
   url = {https://onlinelibrary.wiley.com/doi/full/10.1890/11-0889.1 https://onlinelibrary.wiley.com/doi/abs/10.1890/11-0889.1 https://esajournals.onlinelibrary.wiley.com/doi/10.1890/11-0889.1},
   year = {2012},
}

@book{Brockwell2016,
   author = {Peter J. Brockwell and Richard A. Davis},
   city = {Cham},
   doi = {10.1007/978-3-319-29854-2},
   isbn = {978-3-319-29852-8},
   publisher = {Springer International Publishing},
   title = {Introduction to Time Series and Forecasting},
   url = {http://link.springer.com/10.1007/978-3-319-29854-2},
   year = {2016},
}

@article{Ghadami2021_trafficCongestion,
   author = {Amin Ghadami and Bogdan I. Epureanu},
   doi = {10.1109/TITS.2020.2964021},
   issn = {15580016},
   issue = {2},
   journal = {IEEE Transactions on Intelligent Transportation Systems},
   month = {2},
   pages = {1196-1205},
   publisher = {Institute of Electrical and Electronics Engineers Inc.},
   title = {Forecasting the Onset of Traffic Congestions on Circular Roads},
   volume = {22},
   year = {2021},
}

@inproceedings{Rafi2021_Realtime_LOC_mitigation,
author = {Melvin Rafi and James E. Steck and Animesh Chakravarthy},
title = {Real-time Adaptive Optimal Prediction of Safe Control Spaces and Augmented-Reality Head-Up Displays Towards Aircraft Loss-of-Control Mitigation},
booktitle = {AIAA Scitech 2021 Forum},
year = {2021},
doi = {10.2514/6.2021-0757},
URL = {https://arc.aiaa.org/doi/abs/10.2514/6.2021-0757},
eprint = {https://arc.aiaa.org/doi/pdf/10.2514/6.2021-0757}
}

@article{Zhidong2022_FEP_Reachability,
author = {Lu, Zhidong and Hong, Haichao and Gerdts, Matthias and Holzapfel, Florian},
title = {Flight Envelope Prediction via Optimal Control-Based Reachability Analysis},
journal = {Journal of Guidance, Control, and Dynamics},
volume = {45},
number = {1},
pages = {185-195},
year = {2022},
doi = {10.2514/1.G006219},
URL = { 
        https://doi.org/10.2514/1.G006219
},
eprint = { 
        https://doi.org/10.2514/1.G006219
}
}

@article{Zogopoulos2021_FTC_FixedWingUAV_FE_awareness,
author = {Zogopoulos-Papaliakos, George and Karras, George C. and Kyriakopoulos, Kostas J.},
title = {A Fault-Tolerant Control Scheme for Fixed-Wing {UAV}s with Flight Envelope Awareness},
year = {2021},
issue_date = {Jun 2021},
publisher = {Kluwer Academic Publishers},
address = {USA},
volume = {102},
number = {2},
issn = {0921-0296},
url = {https://doi.org/10.1007/s10846-021-01393-3},
doi = {10.1007/s10846-021-01393-3},
journal = {J. Intell. Robotics Syst.},
month = {jun},
numpages = {33}
}

@inproceedings{Barlow2011_EstimatingLOC_DataDriven,
author = {Jonathan Barlow and Vahram Stepanyan and Kalmanje Krishnakumar},
title = {Estimating Loss-of-Control: A Data-Based Predictive Control Approach},
booktitle = {AIAA Guidance, Navigation, and Control Conference},
year = {2011},
doi = {10.2514/6.2011-6408},
URL = {https://arc.aiaa.org/doi/abs/10.2514/6.2011-6408},
eprint = {https://arc.aiaa.org/doi/pdf/10.2514/6.2011-6408}
}

@inproceedings{bansal2017hamilton,
  title={Hamilton-jacobi reachability: A brief overview and recent advances},
  author={Bansal, Somil and Chen, Mo and Herbert, Sylvia and Tomlin, Claire J},
  booktitle={2017 IEEE 56th Annual Conference on Decision and Control (CDC)},
  pages={2242--2253},
  year={2017},
  organization={IEEE}
}

@article{Altena_ANN_LOC,
author = {Altena, Anique V.N. and van Beers, Jasper J. and de Visser, Coen C.},
title = {Loss-of-Control Prediction of a Quadcopter Using Recurrent Neural Networks},
journal = {Journal of Aerospace Information Systems},
volume = {20},
number = {10},
pages = {648-659},
year = {2023},
doi = {10.2514/1.I011231},
URL = { 
    
        https://doi.org/10.2514/1.I011231
},
eprint = { 
        https://doi.org/10.2514/1.I011231   
}
}

@INPROCEEDINGS{RobotFailure2004,
  author={Carlson, J. and Murphy, R.R. and Nelson, A.},
  booktitle={IEEE International Conference on Robotics and Automation, 2004. Proceedings. ICRA '04. 2004}, 
  title={Follow-up analysis of mobile robot failures}, 
  year={2004},
  volume={5},
  number={},
  pages={4987-4994 Vol.5},
  doi={10.1109/ROBOT.2004.1302508},
  ISSN={1050-4729},
  month={April},
  note ={%[Accessed 13-09-2023]}
}

@article{OLIVER2019772_acLOC,
title = {Safe limits, mindful organizing and loss of control in commercial aviation},
journal = {Safety Science},
volume = {120},
pages = {772-780},
year = {2019},
issn = {0925-7535},
doi = {https://doi.org/10.1016/j.ssci.2019.08.018},
url = {https://www.sciencedirect.com/science/article/pii/S0925753519301055},
author = {Nick Oliver and Thomas Calvard and Kristina Potočnik},
note ={%[Accessed 13-09-2023]}
}

@article{YANG2022105623_robotsafer,
title = {Robot application and occupational injuries: Are robots necessarily safer?},
journal = {Safety Science},
volume = {147},
pages = {105623},
year = {2022},
issn = {0925-7535},
doi = {https://doi.org/10.1016/j.ssci.2021.105623},
url = {https://www.sciencedirect.com/science/article/pii/S092575352100463X},
author = {Siying Yang and Yifan Zhong and Dawei Feng and Rita Yi Man Li and Xue-Feng Shao and Wei Liu}
,note ={%[Accessed 13-09-2023]}
}

@Article{app12031047,
AUTHOR = {Aslan, Muhammet Fatih and Durdu, Akif and Sabanci, Kadir and Ropelewska, Ewa and Gültekin, Seyfettin Sinan},
TITLE = {A Comprehensive Survey of the Recent Studies with {UAV} for Precision Agriculture in Open Fields and Greenhouses},
JOURNAL = {Applied Sciences},
VOLUME = {12},
YEAR = {2022},
NUMBER = {3},
ARTICLE-NUMBER = {1047},
URL = {https://www.mdpi.com/2076-3417/12/3/1047},
ISSN = {2076-3417},
DOI = {10.3390/app12031047},
note ={%[Accessed 13-09-2023]}
}

@article{nisingizwe2022effect,
  title={Effect of unmanned aerial vehicle (drone) delivery on blood product delivery time and wastage in {Rwanda}: A retrospective, cross-sectional study and time series analysis},
  author={Nisingizwe, Marie Paul and Ndishimye, Pacifique and Swaibu, Katare and Nshimiyimana, Ladislas and Karame, Prosper and Dushimiyimana, Valentine and Musabyimana, Jean Pierre and Musanabaganwa, Clarisse and Nsanzimana, Sabin and Law, Michael R},
  journal={The Lancet Global Health},
  volume={10},
  number={4},
  pages={e564--e569},
  year={2022},
  publisher={Elsevier},
  note ={%[Accessed 13-09-2023]}
}

@Article{drones6060137,
AUTHOR = {Nordin, Mohd Hisham and Sharma, Sanjay and Khan, Asiya and Gianni, Mario and Rajendran, Sulakshan and Sutton, Robert},
TITLE = {Collaborative Unmanned Vehicles for Inspection, Maintenance, and Repairs of Offshore Wind Turbines},
JOURNAL = {Drones},
VOLUME = {6},
YEAR = {2022},
NUMBER = {6},
ARTICLE-NUMBER = {137},
URL = {https://www.mdpi.com/2504-446X/6/6/137},
ISSN = {2504-446X},
DOI = {10.3390/drones6060137},
note ={%[Accessed 13-09-2023]}
}

@ARTICLE{Sun_ControlDoubleFailure,
    author={S. {Sun} and X. {Wang} and Q. {Chu} and C. d. {Visser}},
    journal={IEEE Transactions on Robotics}, 
    title={Incremental Nonlinear Fault-Tolerant Control of a Quadrotor With Complete Loss of Two Opposing Rotors}, 
    year={2020},
    volume={},
    number={},
    pages={1-15},
    doi={10.1109/TRO.2020.3010626},
    ISSN={1941-0468},
    month={},
note = {%[Accessed 25-05-2023]}
}

@inproceedings{Sun2019_MonteCarlo,
    author = {S. Sun and C.C. de Visser},
    title = {Quadrotor Safe Flight Envelope Prediction in the High-Speed Regime: A Monte-Carlo Approach},
    booktitle = {AIAA Scitech 2019 Forum},
    year = {2019},
    doi = {10.2514/6.2019-0948},
    URL = {https://arc.aiaa.org/doi/abs/10.2514/6.2019-0948},
    publisher = {American Institute of Aeronautics and Astronautics},
    note ={%[Accessed 15-09-2023]}
}

@ARTICLE{Ke_2023_Quadrotor_123FTC,
  author={Ke, Chenxu and Cai, Kai-Yuan and Quan, Quan},
  journal={IEEE Transactions on Robotics}, 
  title={Uniform Passive Fault-Tolerant Control of a Quadcopter With One, Two, or Three Rotor Failure}, 
  year={2023},
  volume={39},
  number={6},
  pages={4297-4311},
  doi={10.1109/TRO.2023.3297048}}

@ARTICLE{Guo2022_Quad_GE_blade_damage,
  author={Guo, Kexin and Zhang, Wenyu and Zhu, Yukai and Jia, Jindou and Yu, Xiang and Zhang, Youmin},
  journal={IEEE Transactions on Industrial Electronics}, 
  title={Safety Control for Quadrotor {UAV} Against Ground Effect and Blade Damage}, 
  year={2022},
  volume={69},
  number={12},
  pages={13373-13383},
  doi={10.1109/TIE.2022.3140494}}

@ARTICLE{Nan2022_NMPC_FTC,
  author={Nan, Fang and Sun, Sihao and Foehn, Philipp and Scaramuzza, Davide},
  journal={IEEE Robotics and Automation Letters}, 
  title={Nonlinear {MPC} for Quadrotor Fault-Tolerant Control}, 
  year={2022},
  volume={7},
  number={2},
  pages={5047-5054},
  doi={10.1109/LRA.2022.3154033}}

@ARTICLE{Eltrabyly2022_QuadFTC,
  author={Eltrabyly, Akram and Ichalal, Dalil and Mammar, Said},
  journal={IEEE Control Systems Letters}, 
  title={Quadcopter Trajectory Tracking in the Presence of 4 Faulty Actuators: A Nonlinear {MHE} and {MPC} Approach}, 
  year={2022},
  volume={6},
  number={},
  pages={2024-2029},
  doi={10.1109/LCSYS.2021.3137099}}

@ARTICLE{Gurriet2020_SCCF_nonlinearsys,
  author={Gurriet, Thomas and Mote, Mark and Singletary, Andrew and Nilsson, Petter and Feron, Eric and Ames, Aaron D.},
  journal={IEEE Access}, 
  title={A Scalable Safety Critical Control Framework for Nonlinear Systems}, 
  year={2020},
  volume={8},
  number={},
  pages={187249-187275},
  doi={10.1109/ACCESS.2020.3025248}}

@ARTICLE{Singletary2022_OnboardSafety_RacingQuad_GeofencingCBF,
  author={Singletary, Andrew and Swann, Aiden and Chen, Yuxiao and Ames, Aaron D.},
  journal={IEEE Robotics and Automation Letters}, 
  title={Onboard Safety Guarantees for Racing Drones: High-Speed Geofencing With Control Barrier Functions}, 
  year={2022},
  volume={7},
  number={2},
  pages={2897-2904},
  doi={10.1109/LRA.2022.3144777}}

@INPROCEEDINGS{Singletary2022_SafeDroneFlight_TBC,
  author={Singletary, Andrew and Swann, Aiden and Rodriguez, Ivan Dario Jimenez and Ames, Aaron D.},
  booktitle={2022 IEEE/RSJ International Conference on Intelligent Robots and Systems (IROS)}, 
  title={Safe Drone Flight with Time-Varying Backup Controllers}, 
  year={2022},
  volume={},
  number={},
  pages={4577-4584},
  doi={10.1109/IROS47612.2022.9981741}}

@ARTICLE{Zheng2023_Geofencing_Quad,
  author={Zheng, Zhi and Su, Xiaojie and Jiang, Tao and Huang, Jiangshuai},
  journal={IEEE Transactions on Industrial Electronics}, 
  title={Robust Dynamic Geofencing Attitude Control for Quadrotor Systems}, 
  year={2023},
  volume={70},
  number={2},
  pages={1861-1869},
  doi={10.1109/TIE.2022.3159919}}

@ARTICLE{Ghadami2022_TrafficJam,
  author={Ghadami, Amin and Doering, Charles R. and Drake, John M. and Rohani, Pejman and Epureanu, Bogdan I.},
  journal={IEEE Transactions on Intelligent Transportation Systems}, 
  title={Stability and Resilience of Transportation Systems: Is a Traffic Jam About to Occur?}, 
  year={2022},
  volume={23},
  number={8},
  pages={10803-10814},
  doi={10.1109/TITS.2021.3095897}}

@article{tredennick2022anticipating_measles,
  title={Anticipating infectious disease re-emergence and elimination: A test of early warning signals using empirically based models},
  author={Tredennick, Andrew T and O’Dea, Eamon B and Ferrari, Matthew J and Park, Andrew W and Rohani, Pejman and Drake, John M},
  journal={Journal of the Royal Society Interface},
  volume={19},
  number={193},
  pages={20220123},
  year={2022},
  publisher={The Royal Society}
}

@article{ditlevsen2023warning_AMOC_CSD,
  title={Warning of a forthcoming collapse of the Atlantic meridional overturning circulation},
  author={Ditlevsen, Peter and Ditlevsen, Susanne},
  journal={Nature Communications},
  volume={14},
  number={1},
  pages={1--12},
  year={2023},
  publisher={Nature Publishing Group}
}

@article{Fridovich-Keil2020_CollisionAvoidance,
author = {David Fridovich-Keil and Andrea Bajcsy and Jaime F Fisac and Sylvia L Herbert and Steven Wang and Anca D Dragan and Claire J Tomlin},
title ={Confidence-aware motion prediction for real-time collision avoidance},
journal = {The International Journal of Robotics Research},
volume = {39},
number = {2-3},
pages = {250-265},
year = {2020},
doi = {10.1177/0278364919859436},
URL = { 
        https://doi.org/10.1177/0278364919859436
},
eprint = { 
        https://doi.org/10.1177/0278364919859436}
,
}

@ARTICLE{Madruga2023_LOE_UAV_actuators,
  author={Madruga, Sarah P. and Nascimento, Tiago P. and Holzapfel, Florian and Lima, Antonio M. N.},
  journal={IEEE Robotics and Automation Letters}, 
  title={Estimating the Loss of Effectiveness of {UAV} Actuators in the Presence of Aerodynamic Effects}, 
  year={2023},
  volume={8},
  number={3},
  pages={1335-1342},
  doi={10.1109/LRA.2023.3238184}}

@article{forzieri2022emerging,
  title={Emerging signals of declining forest resilience under climate change},
  author={Forzieri, Giovanni and Dakos, Vasilis and McDowell, Nate G and Ramdane, Alkama and Cescatti, Alessandro},
  journal={Nature},
  volume={608},
  number={7923},
  pages={534--539},
  year={2022},
  publisher={Nature Publishing Group UK London}
}

@article{kerr2023haptic,
  title={Identifying critical transitions and instability in haptic systems},
  author={Kerr, Liam and Hutchison, Chantal and K{\"o}vecses, J{\'o}zsef},
  journal={Nonlinear Dynamics},
  volume={111},
  number={13},
  pages={12607--12623},
  year={2023},
  publisher={Springer}
}

@article{
Scheffer_2008_CriticalTransitions,
author = {Marten Scheffer  and Stephen R. Carpenter  and Timothy M. Lenton  and Jordi Bascompte  and William Brock  and Vasilis Dakos  and Johan van de Koppel  and Ingrid A. van de Leemput  and Simon A. Levin  and Egbert H. van Nes  and Mercedes Pascual  and John Vandermeer },
title = {Anticipating Critical Transitions},
journal = {Science},
volume = {338},
number = {6105},
pages = {344-348},
year = {2012},
doi = {10.1126/science.1225244},
URL = {https://www.science.org/doi/abs/10.1126/science.1225244},
eprint = {https://www.science.org/doi/pdf/10.1126/science.1225244}}

@INPROCEEDINGS{vanBeers_FCM,
  author={Van Beers, Jasper J. and Solanki, Prashant and De Visser, Coen C.},
  booktitle={2024 IEEE International Conference on Robotics and Automation (ICRA)}, 
  title={A novel metric for detecting quadrotor loss-of-control}, 
  year={2024},
  volume={},
  number={},
  pages={15570-15576},
  doi={10.1109/ICRA57147.2024.10610662}}

@Article{drones9020141_flyaways,
AUTHOR = {Farajijalal, Mohsen and Eslamiat, Hossein and Avineni, Vikrant and Hettel, Eric and Lindsay, Clark},
TITLE = {Safety Systems for Emergency Landing of Civilian Unmanned Aerial Vehicles ({UAV}s)—A Comprehensive Review},
JOURNAL = {Drones},
VOLUME = {9},
YEAR = {2025},
NUMBER = {2},
ARTICLE-NUMBER = {141},
URL = {https://www.mdpi.com/2504-446X/9/2/141},
ISSN = {2504-446X},
DOI = {10.3390/drones9020141}
}

@article{Li2024,
   author = {Tinghua Li and Bayu Jayawardhana},
   journal = {IEEE Transactions on Automatic Control},
   publisher = {IEEE},
   title = {Collision-free source seeking control methods for unicycle robots},
   year = {2024}
}

@ARTICLE{Fang2020_FDDAutonomousVehicles,
  author={Fang, Yukun and Min, Haigen and Wang, Wuqi and Xu, Zhigang and Zhao, Xiangmo},
  journal={IEEE Sensors Journal}, 
  title={A Fault Detection and Diagnosis System for Autonomous Vehicles Based on Hybrid Approaches}, 
  year={2020},
  volume={20},
  number={16},
  pages={9359-9371},
  doi={10.1109/JSEN.2020.2987841}}

@article{vanBeers2026_ews_for_loc,
author = {Jasper J. van Beers  and Marten Scheffer  and Prashant Solanki  and Ingrid A. van de Leemput  and Egbert H. van Nes  and Coen C. de Visser },
title = {Early warning signals for loss of control in complex systems},
journal = {Proceedings of the National Academy of Sciences},
volume = {123},
number = {27},
pages = {e2608847123},
year = {2026},
doi = {10.1073/pnas.2608847123},
URL = {https://www.pnas.org/doi/abs/10.1073/pnas.2608847123},
}

\end{document}